\documentclass[letterpaper,twocolumn,10pt]{article}
\usepackage{usenix-2020-09}

\usepackage{url}
\usepackage{subfig}
\usepackage{multirow}
\usepackage{tikz}
\usetikzlibrary{positioning}
\usepackage{comment}
\usepackage{graphicx}
\usepackage{xspace}
\usepackage{xcolor}
\usepackage{amsmath}
\usepackage{cleveref}
\usepackage{hyperref}
\usepackage{booktabs}
\usepackage{amssymb}
\usepackage{enumitem}

\begin{document}

\newcommand{\sysname}{CCAaaS\xspace}

\title{\sysname: An Orchestration Platform for High-Velocity Design, Development, and Deployment of Data Center Transports at Meta}
\newcommand{\FB}{Meta\xspace}
\newcommand{\CCALogger}{CCALogger\xspace}
\newcommand{\DTCP}{DCNDelay-CCA\xspace}
\newcommand{\Glide}{XRegion-CCA\xspace}

\newcommand{\BPF}{eBPF\xspace}
\newcommand{\eBPF}{eBPF\xspace}
\newcommand{\bpf}{eBPF\xspace}
\newcommand{\ebpf}{eBPF\xspace}

\newcommand{\program}{eBPF program\xspace}
\newcommand{\programs}{eBPF programs\xspace}

\makeatletter
\renewcommand\section{\@startsection {section}{1}{\z@}%
                                   {-2.5ex plus -1ex minus -.2ex}%
                                   {1.5ex plus .2ex}%
                                   {\reset@font\large\bf}}
\renewcommand\subsection{\@startsection{subsection}{2}{\z@}%
                                     {-2.0ex plus -0.5ex minus -.2ex}%
                                     {1.0ex plus .2ex}%
                                     {\reset@font\normalsize\bf}}
\renewcommand\subsubsection{\@startsection{subsubsection}{3}{\z@}%
                                     {-1.5ex plus -0.5ex minus -.2ex}%
                                     {0.8ex plus .2ex}%
                                     {\reset@font\normalsize\bf}}
\makeatother

\setlength{\textfloatsep}{8pt plus 2pt minus 2pt}
\setlength{\floatsep}{8pt plus 2pt minus 2pt}
\setlength{\intextsep}{8pt plus 2pt minus 2pt}
\setlength{\dbltextfloatsep}{8pt plus 2pt minus 2pt}
\setlength{\dblfloatsep}{8pt plus 2pt minus 2pt}

\setlength{\abovecaptionskip}{6pt plus 1pt minus 1pt}
\setlength{\belowcaptionskip}{0pt}

\setlength{\abovedisplayskip}{6pt plus 2pt minus 2pt}
\setlength{\belowdisplayskip}{6pt plus 2pt minus 2pt}

\newlength{\subsecvspace}
\setlength{\subsecvspace}{-3pt} 

\author{
{Prankur Gupta}\\
{Meta}
\and 
{Balasubramanian Madhavan}\\
{Meta}
\and 
{Miao Xu}\\
{Meta}
\and 
{Prashanth Kannan}\\
{Meta}
\and 
{Jie Meng}\\
{Meta}
\and 
{Neil Spring}\\
{Meta}
\and 
{Rajiv Krishnamurthy}\\
{Meta}
\and 
{Srikanth Sundaresan}\\
{Meta}
\and 
{Martin Lau}\\
{Meta}
\and 
{Theophilus A. Benson}\\
{Carnegie Mellon University}
}
\maketitle
\pagestyle{empty}
\thispagestyle{empty}

\begin{abstract}
Today, data center networks run multiple congestion control algorithms (CCAs), each tailored to distinct traffic characteristics, from latency-sensitive intra-region RPCs to high-throughput cross-region AI workloads. The rapid evolution of AI training and inference further accelerates this trend by continually introducing new traffic characteristics that demand new CCAs.

This paper introduces CCA-as-a-Service (\sysname), our eBPF-based ecosystem for the rapid design, deployment, and operation of production-grade CCAs at hyperscale. We present (1) a modular BPF-TCP framework and general lessons for designing modular CCAs; (2) a deployment and orchestration layer that provides disruption-free CCA updates, along with best practices for managing a zoo of CCAs at scale; and (3) a lightweight observability tool for CCAs and a blueprint for scaling CCA observability.
\sysname manages over 97\% of our server-to-server traffic using three active eBPF-based CCAs; two of the five CCAs we have built have since been deprecated. We show that our eBPF-centric approach to CCAs development delivers the same application-level performance (retransmissions, bandwidth, and tail latency) as their native in-kernel counterparts, and that our custom CCAs add negligible CPU and memory cost over their eBPF baselines. In terms of realized benefits, the CCA built with \sysname significantly improved network, and application metrics, for example, one of our CCAs reduced  P99 latency for cross-region services by up to 20$\times$.

\end{abstract}

\section{Introduction}
\vspace{\subsecvspace}

Large-scale data-center networks must simultaneously satisfy competing objectives, including low latency, high throughput, and high utilization, for a constantly evolving set of applications~\cite{google:bolt,google:swift,metadctcp:nsdi24}. For example, latency-sensitive RPCs for end-user services (e.g., News Feed, Ads ranking) demand predictable tail latency~\cite{google:bolt}, while AI training workloads (e.g., checkpointing and large-scale data replication) require high throughput~\cite{meta:ebb}. In such environments with competing objectives, a single congestion control algorithm (CCA) is insufficient to meet all requirements.

\begin{figure}[t]
    \centering
    \includegraphics[width=0.75\linewidth]{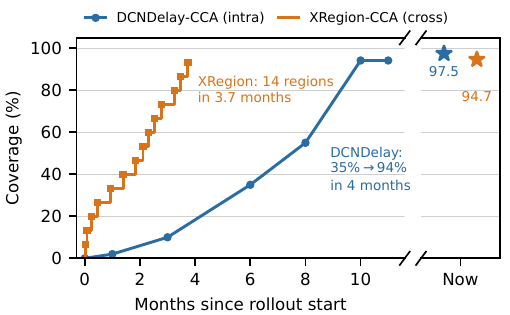}
     \caption{Rollout timelines for our two most recent CCAs: \DTCP is rolled out per region, while \Glide is rolled out per region.}

    \label{fig:rollout}
\end{figure}

As a result, modern hyperscaler networks simultaneously deploy multiple CCAs, each optimized for a specific class of infrastructure, class of traffic, or application characteristics. While prior work has described the deployment of individual CCAs for varying objectives~\cite{metadctcp:nsdi24,dctcp:morganstanley,google:bolt,google:swift,google:timely,falcon:google,dcqcn,bytedance:barre}, in practice, many of these operators actively manage a growing portfolio of CCAs that coexist within the same infrastructure and must evolve independently over time to address shifting priorities and emerging use cases. Yet, few papers have discussed the systems challenges of developing, operating, and evolving a heterogeneous set of CCAs in production.
We believe discussing these challenges is important because we see, in our infrastructure, the trend of rapid CCA development and deployment being accelerated by AI workloads, which rapidly introduce new traffic patterns and use cases that significantly shortens the lifecycle from CCA conception to production deployment (\cref{fig:rollout}).

In this paper, we present \sysname, a culmination of our experience  pursuing in-kernel programmability via eBPF, which shows that operating CCAs raises challenges and aspect of CCA design generally overlooked by recent eBPF-based TCP CCA frameworks~\cite{tran2019beyond,etran}: rollout safety, rollback correctness, debugging, and observability. We note that more general production CCA work focus on individual CCA design and not the challenges of managing co-existence~\cite{metadctcp:nsdi24,google:bolt}.

Our experience designing \sysname  has highlighted four common challenges:

\begin{table}[t]
\centering
\scriptsize
\begin{tabular}{l l l l}
\toprule
\textbf{CCA} & \textbf{Dev. Period} & \textbf{Deploy Period} & \textbf{Status} \\
\midrule
Best Effort CC & 6 months & 1--2 years & Deprecated \\

BPF-DCTCP~\cite{metadctcp:nsdi24}  & 6 months & 1 year & Deprecated \\
\DTCP (\cref{sec:modular:d2tcp}) & 9 months & 10 months & Active \\
\Glide (\cref{sec:modular:xregion})  & 6 months & 6 months & Active  \\
BPF-Cubic  & 4.5 months & 3 months & Active$^{\dagger}$ \\
\bottomrule
\end{tabular}
\caption{CCAs developed and deployed at Meta using the \sysname framework. $^{\dagger}$At the time of writing, BPF-Cubic was still deployed, but we were actively reducing its usage as \Glide supersedes it.} 
\label{tab:cca_timeline}
\end{table}
\begin{itemize}[leftmargin=*, nosep]
\item
\textbf{C1, Rapid iteration under tight performance constraints:} CCA development demands fast experimentation, but congestion control executes on the critical datapath, where even small abstraction overhead is unacceptable.
\item\textbf{C2, Safe and incremental rollout:} Certain CCAs require both endpoints to be upgraded (e.g., for ECN negotiation), which complicates rollout. Moreover, new logic must apply retroactively to existing connections, since long-lived flows (e.g., AI training) would otherwise run stale CCA logic until they are reset.
\item\textbf{C3, Timely and correct rollback:} Rollback is complicated by sticky transport state: once ECN is negotiated, it persists even after the CCA changes, and fallback CCAs may misinterpret it.
\item\textbf{C4, Lightweight, CCA-aware observability:} Operators need visibility into CCA-specific internal state to validate changes and diagnose incidents, but existing tools are too heavyweight (e.g., \texttt{tcpdump})  or too rigid (e.g.,  Web10G~\cite{web10g_project}).
\end{itemize}

In this paper, we describe our experience building and operating \emph{CCA-as-a-Service (\sysname)}, an eBPF-based ecosystem that enables rapid development, safe evolution and deployment of CCAs coexisting atop the same production infrastructure. With \sysname, developers design CCA by implementing programs specified by BPF-TCP hooks, and the ecosystem handles its orchestration (e.g., , rollout, rollback) and monitoring. Our design of \sysname builds on three insights:  First, to ensure efficiency and performance,  CCA interfaces should be specific rather than generic  and allow control over various layers to the network stack  beyond CCA core control loop. Second, CCA deployment (or rollback on failures) must explicitly account for transport related state or functionality provided by the network, such as negotiated ECN. Third, to scalably observe CCAs,  observability should be co-designed with CCA control loop behavior and codebase. \sysname builds on NetEdit~\cite{meta:netedit}, our production eBPF orchestrator, and extends it to provide CCA-specific support (e.g., CCA-specific state sharing).

\sysname has been in production for over five years, has been used to develop five CCAs (\cref{tab:cca_timeline}), and currently manages over 97\% of server-to-server traffic at Meta through custom eBPF-based CCAs. Our two most recent CCAs provided significant production wins: \DTCP reduced retransmissions for major services by 30--40\%, while \Glide, reduced cross-region service P99 latency by up to 20$\times$.  We also identify several key lessons for CCA developers and infrastructure administrator: our lessons range from CCA management to eBPF orchestration.

This paper makes the following contributions:
\begin{itemize}[leftmargin=*, nosep]
\item \textbf{Modular BPF-TCP framework (C1):} We introduce an eBPF-based framework for developing modular CCAs and evolve it to provide control over other important functions and configurations within the kernel networking stack (\cref{sec:modular})
\item \textbf{CCA deployment and orchestration ecosystem (C2 and C3):} We extend an existing production-grade eBPF orchestration system to support CCAs, introducing primitives for safe rollback and coordination primitives for safe, disruption-free rollout (\cref{sec:ecosystem}). 
\item \textbf{\CCALogger (C4):} We provide the first lightweight and comprehensive CCA observability framework, which scales by co-designing its observability primitives with BPF-TCP's modular decomposition and lifecycle (\cref{sec:ccalogger}). 
\item \textbf{Deployment lessons:} We distill a set of lessons from the challenges we have faced in designing CCAs, managing CCAs, and evolving eBPF to support an ecosystem of eBPF-based CCAs (\cref{sec:operational}).
\end{itemize}

\section{Motivation}
\label{sec:motivation}
\vspace{\subsecvspace}
In this section, we describe our production environment to highlight why we run multiple CCAs (\cref{sec:motivation:portfolio}) and to justify our choice of eBPF for building our CCAs (\cref{sec:motivation:ebpf}). The data in this section were collected across our fleet via FBFlow and queries to our infrastructure database, and span three months in 2026.
\subsection{The Production Case for Multiple Simultaneous CCAs}
\label{sec:motivation:portfolio}
\vspace{\subsecvspace}

\begin{figure}[t]
    \centering
    \includegraphics[width=0.75\linewidth]{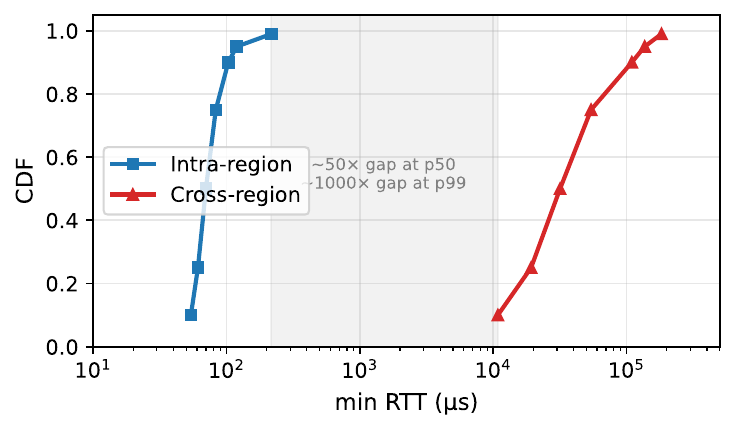}
    \caption{CDF of minimum RTT by traffic class: intra-region versus cross-region flows.}
    \label{fig:rtt-cdf}
    \vspace{-8pt}
\end{figure}

\textbf{Heterogeneous Traffic Conditions.} Modern hyperscale networks must support applications with fundamentally different traffic characteristics and requirements. \Cref{fig:rtt-cdf} illustrates the gap between the RTTs of intra-region flows (median ${\sim}$70\,$\mu$s) and cross-region flows (median ${\sim}$32\,ms). This difference of nearly three orders of magnitude translates into distinct CCA requirements, and it is precisely why we run multiple CCAs across our fleet (\cref{tab:cca-metrics}).

\begin{figure}[t]
    \centering
    \includegraphics[width=0.75\linewidth]{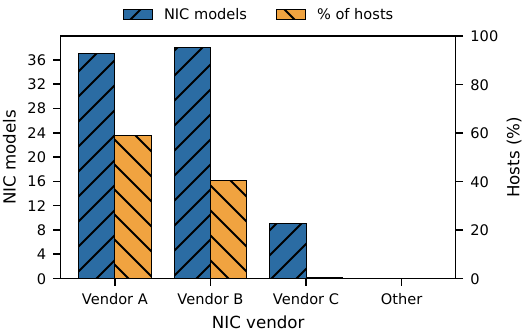}
    \caption{Top-3 NIC vendors by deployment fraction. We use a predefined OCP NIC configuration, but different vendors supply the components; a model is a set of NICs with the same vendor and hardware combination.}
    \label{fig:vendors_models_hosts}
\end{figure}

\begin{figure}[t]
    \centering
    \includegraphics[width=0.75\linewidth]{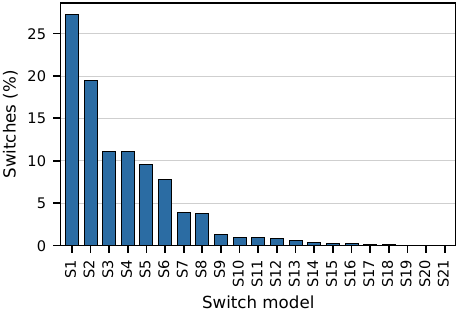}
    \caption{Distribution of switches deployed at Meta.}
    \label{fig:switch_model_pct}
\end{figure}

\textbf{Drawbacks of ECN for Intra-Region Traffic.}
We recently showed that within a region, congestion (and loss) generally occurs at end hosts and in NIC buffers rather than in network switches~\cite{millisampler:imc22}. However, only switches generate ECN marks, so existing ECN-based CCAs cannot accurately capture this congestion. Additionally, our data centers comprise multiple hardware generations across NIC vendors and switch platforms (\cref{fig:vendors_models_hosts,fig:switch_model_pct}), and each generation introduces different queuing and ECN behavior into our network. Thus, selecting ECN configurations that effectively support both incast workloads and bulk transfers is challenging. Because of this configuration complexity and ECN's limited visibility into host-side congestion, we opt for a delay-based CCA for intra-region traffic (\cref{sec:modular:d2tcp}).

\textbf{The Shift Toward ECN for Cross-Region Traffic.} Although we planned to move intra-region traffic to a delay-based CCA, our experience with delay on cross-region traffic showed that delay alone was an insufficient signal: changes in cross-region delay stem either from congestion on cross-region links or from our traffic-engineering system rerouting traffic. We also observed that loss-driven CCAs, e.g., Cubic, perform poorly for our long-lived AI workloads because they react to congestion too late, an observation shared by others. Thus, while we use delay for intra-region traffic, we design our cross-region CCA around ECN.

\begin{figure}[t]
    \centering
    \begin{minipage}{0.48\linewidth}
        \centering
        \includegraphics[width=\linewidth]{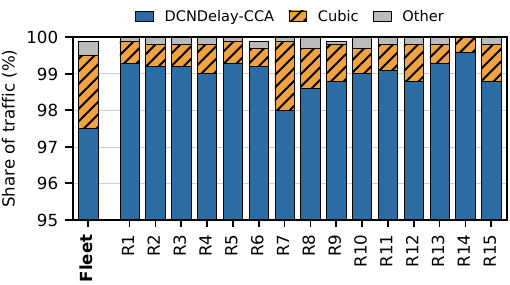}
        \vspace{0.3em}
        \textbf{(a)} Intra-region traffic (\DTCP).
        \label{fig:CCA:deployment:intra}
    \end{minipage}
~
    \begin{minipage}{0.48\linewidth}
        \centering
        \includegraphics[width=\linewidth]{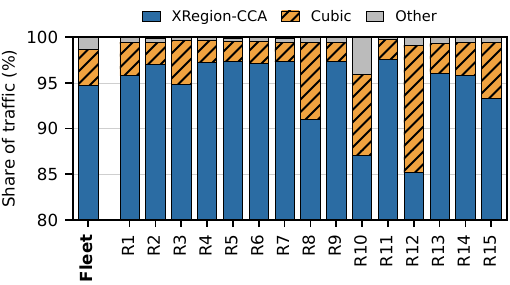}
        \vspace{0.3em}
        \textbf{(b)} Cross-region traffic (\Glide).
        \label{fig:CCA:deployment:cross}
    \end{minipage}

    \caption{Per-region distribution of CCA usage for server-to-server traffic (by bytes/s). Note the truncated y-axes.}
    \label{fig:CCA:deployment}
\end{figure}

\textbf{The Need for Host-Level CCA Coexistence.} Although we design our CCAs with different objectives, intra-region and cross-region flows must coexist on the same hosts, NICs, and rack switches. Consequently, aggressive behavior by one class of flows may starve or otherwise harm the other. Moreover, each CCA deployment introduces constraints that limit future options. For example, when we first deployed DCTCP, we could easily tailor ECN marking configurations to it; however, these settings became difficult to maintain once we introduced other CCAs. \Cref{tab:cca-metrics,fig:CCA:deployment} show the current distribution of CCAs across our fleet. While our two custom CCAs (\Glide and \DTCP) dominate at the fleet level, usage varies across regions, particularly for cross-region traffic, \Glide requires a kernel upgrade and we are unable to upgrade a subset of hosts and so they still use BPF-Cubic. Managing this portfolio poses two challenges: designing algorithms that coexist safely, and choosing deployment constraints and configurations that leave room for future, unanticipated CCAs.

\begin{table}[t]
\centering
\scriptsize
\begin{tabular}{lccccc}
\toprule
\textbf{Scope} & \textbf{BBR} & \textbf{BPF-Cubic} & \textbf{\Glide} & \textbf{\DTCP} & \textbf{Unclass.} \\
\midrule
All traffic   & $<$0.1\% & 2.2\% & 11.7\% & 85.6\% & 0.5\% \\
Intra-region  & $<$0.1\% & 1.9\% & $<$0.1\% & 97.5\% & 0.4\% \\
Intra-DC      & $<$0.1\% & 4.5\% & $<$0.1\% & 95.2\% & 0.3\% \\
Cross-region  & 0.1\%    & 3.9\% & 94.7\% & $<$0.1\% & 1.2\% \\
\bottomrule
\end{tabular}
\caption{Fleet-wide distribution of CCA usage for server-to-server traffic (by bytes/s). Over 97\% of traffic uses custom eBPF-based CCAs.}
\label{tab:cca-metrics}
\end{table}

\subsection{Why eBPF for Congestion Control}
\label{sec:motivation:ebpf}
\vspace{\subsecvspace}

These challenges call for a mechanism for building CCAs that matches the performance of existing in-kernel CCAs while being flexible enough to deploy without disrupting our applications.

\textbf{Kernel Modifications.} Given our long history of upstreaming kernel patches, an obvious choice is to modify the kernel directly. However, while this approach provides full flexibility, it is operationally infeasible at our iteration speed: kernel patches are slow to merge, and deploying them requires host reboots that disrupt traffic.

\textbf{Kernel Modules.} Although kernel modules are less disruptive than direct kernel modifications, they have two key drawbacks. First, they are tied to the kernel version they are built for, which complicates kernel upgrades. Second, in certain scenarios, replacing them requires rebooting hosts.

\textbf{Userspace Stacks.} Others have shown that moving the network stack, including the CCA, to userspace~\cite{snap,luna} enables rapid iteration; however, this comes at the cost of rewriting the network stack and often requires specialized hardware (e.g., custom NICs). Moreover, these userspace stacks often sacrifice the maturity and efficiency of the kernel's TCP implementation.

\textbf{eBPF} occupies a sweet spot in this design space: eBPF programs execute in the kernel, and thus provide kernel-level performance, yet they can be loaded and replaced at runtime without reboots. Moreover, the eBPF verifier enforces checks that prevent a broad class of safety issues, such as unbounded loops and invalid memory accesses. Because of this combination of kernel-level performance and continuous delivery, we chose eBPF as the substrate for our CCAs. Since making this choice, we have reimplemented Cubic and DCTCP in eBPF and created new bespoke CCAs (\cref{tab:cca_timeline}).

\section{\sysname Overview}
\label{sec:overview}
\vspace{\subsecvspace}

\begin{figure}[t]
    \centering
    \includegraphics[width=0.95\linewidth]{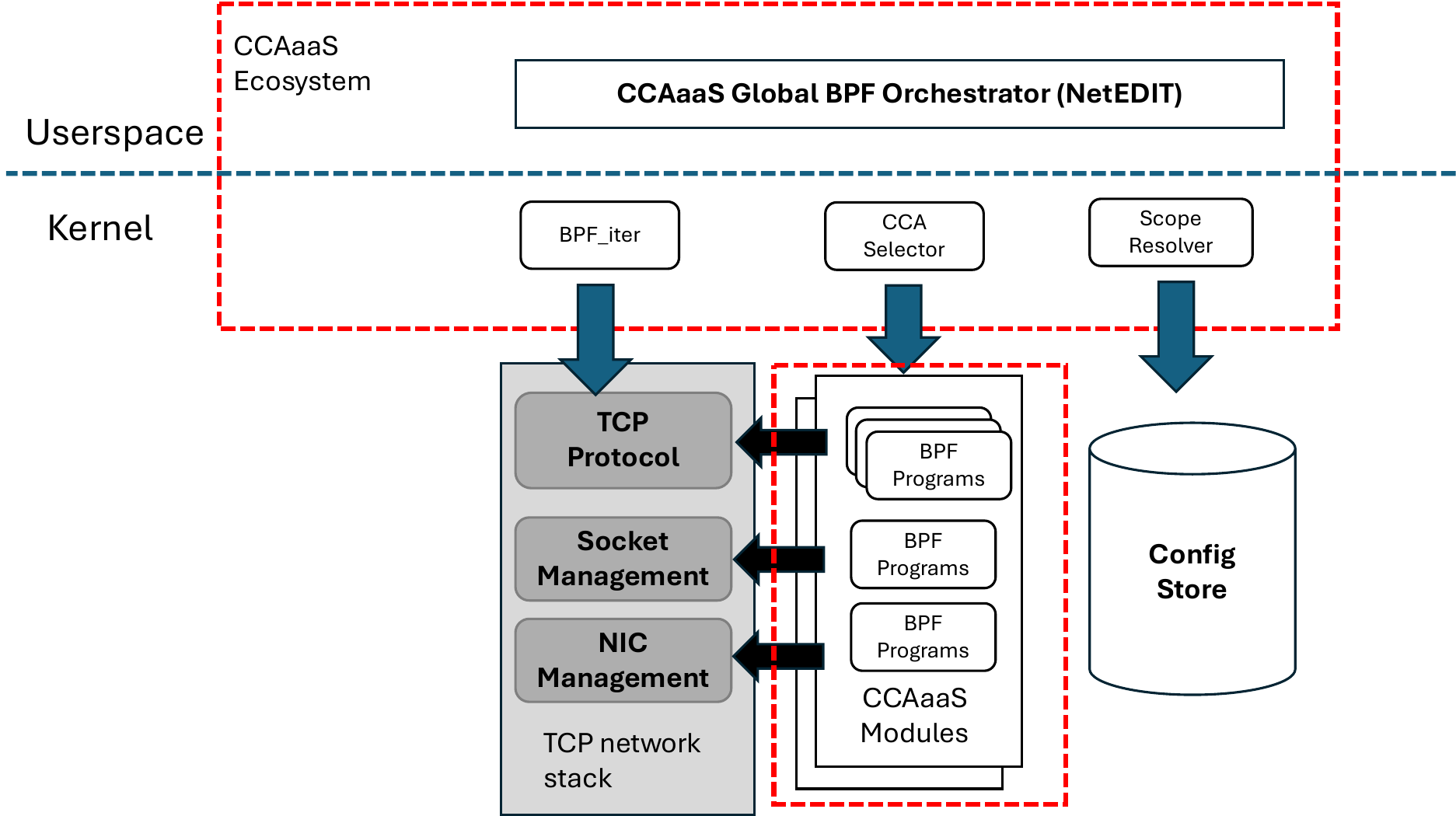}
    \caption{\sysname architecture. }
    \label{fig:ccaaas:architecture}
\end{figure}

Our philosophy is to treat CCA development much like microservice development~\cite{meta:testing}: modular, rapidly developed, and continuously deployed. The \sysname ecosystem (\cref{fig:ccaaas:architecture}) is explicitly designed to support this philosophy and includes the following components: (1) a modular BPF-TCP framework for composing CCAs from reusable eBPF hookpoints (\cref{sec:modular}); (2) a deployment and orchestration layer for disruption-free rollout and rollback (\cref{sec:ecosystem}); and (3) a lightweight observability tool for CCA-specific introspection (\cref{sec:ccalogger}).
While \sysname leverages NetEdit~\cite{meta:netedit} for eBPF program loading and general lifecycle management, in this paper we extend NetEdit with additional CCA-specific management primitives. \sysname splits these components across a control plane and a data plane:

\textbf{Control Plane.} The control plane is never involved in per-packet or per-event processing. Instead, it manages the lifecycle of the eBPF objects that a CCA requires to operate correctly. To do this, the control plane identifies which eBPF programs to deploy, installs them (and their related objects) in the kernel, and determines when to safely uninstall and remove them. The control plane also manages the policies that govern when and how a CCA behaves for a specific connection; it installs these policies in eBPF maps for the eBPF programs to use.

\textbf{Data Plane.} The data plane consists of the eBPF objects, e.g., maps and programs, that are involved in per-packet and per-event processing.

\section{Modular CCA Framework}
\label{sec:modular}
\vspace{\subsecvspace}

Our experience with CCAs over the last five years has revealed that production-grade CCAs require control over the core control loop (e.g., slow start) and, more importantly, over cross-layer kernel functionality (e.g., TSO segmentation) and internal kernel configurations. Our experience has also yielded design principles for efficiently sharing state within a CCA and between CCAs and other eBPF programs.
The current BPF-TCP hookpoint interface (i.e., \texttt{tcp\_congestion\_ops}) is the culmination of the lessons we learned as we refactored, evolved, and enhanced it to support our use cases.

This section describes the BPF-TCP hook interface and its evolution (\cref{sec:modular:hooks}), our state-sharing designs (\cref{sec:modular:sharing}), and two CCAs that we recently deployed with \sysname (\cref{sec:modular:usecases}).

\subsection{The BPF-TCP Hook Interface}
\label{sec:modular:hooks}

Linux's BPF-TCP hookpoint, \texttt{tcp\_congestion\_ops}, is implemented as a \texttt{struct\_ops}, which bundles a set of eBPF programs. This \texttt{struct\_ops} provides a way to implement the kernel's underlying CCA interface. Recall that this interface is a struct of functions: developers implement a CCA by implementing the functions exposed in the struct. BPF-TCP lets developers write a set of eBPF programs, each implementing one of these functions; \cref{tab:cca-hooks} lists the hookpoints that BPF-TCP exposes. We chose to design BPF-TCP as a bundle of eBPF programs within a \texttt{struct\_ops}, rather than as independently defined BPF programs, to ensure that the different functions are always deployed and updated consistently, which is important when CCA code is constantly changing: \texttt{struct\_ops} atomically replaces all callbacks as a unit. Unlike other attach types~\cite{BPFComposer,meta:netedit} (e.g., sockops, TC, or XDP), which allow many BPF programs to operate on the same event, BPF-TCP allows only one \texttt{struct\_ops} (i.e., the programs bundled within it) to handle a connection's packets or events. Thus, there is no need for ordering, prioritization, or conflict-resolution mechanisms among CCAs applied to a connection (or socket), because only one is applied.

\subsubsection{Defining a New CCA}
\begin{figure}[t]
    \centering
    \includegraphics[width=0.55\linewidth]{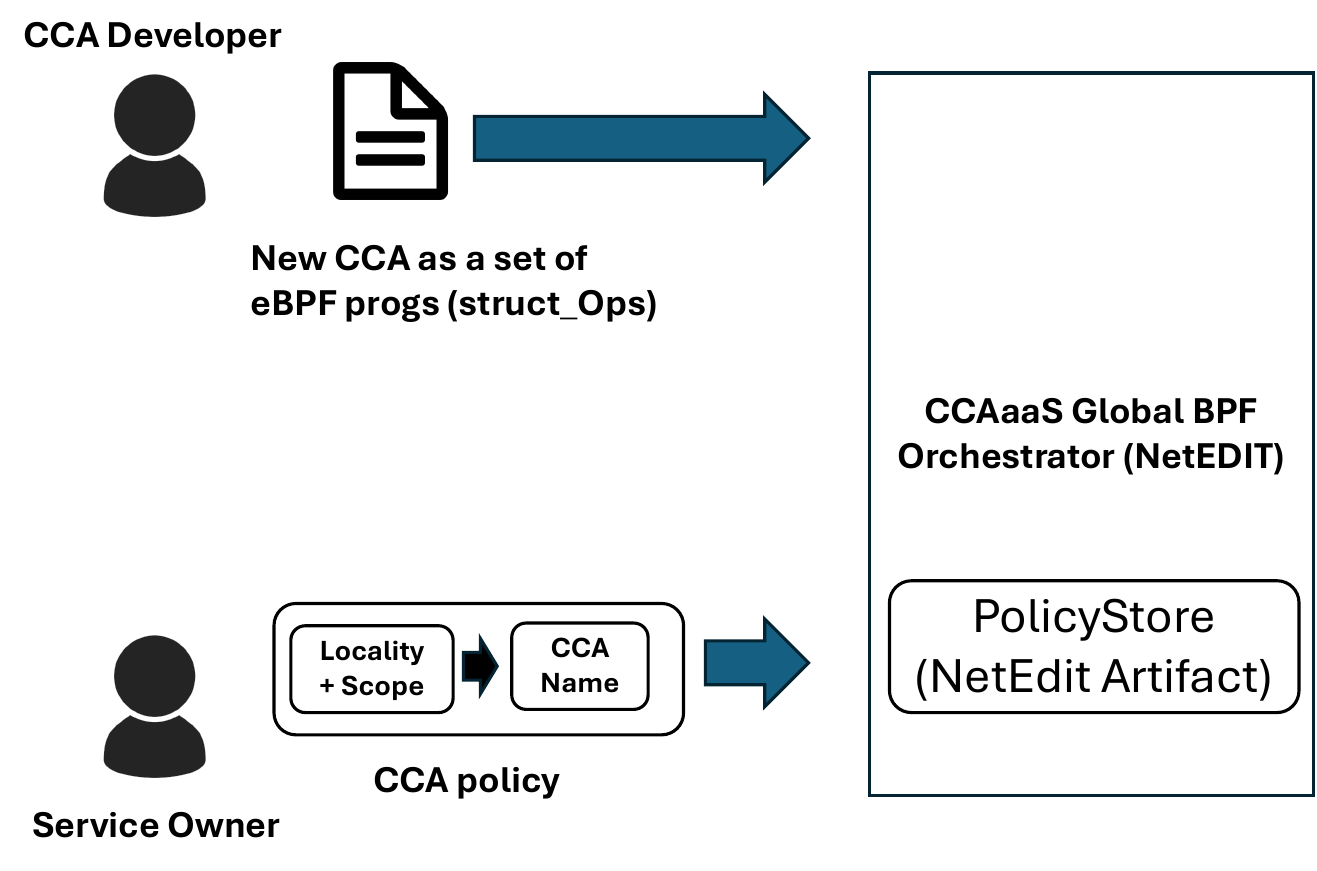}
    \caption{Workflow for defining a new CCA. }
    \label{fig:newcca}
\end{figure}

To add a new CCA to \sysname (\cref{fig:newcca}), (1) the developer implements the new CCA's logic as a set of BPF programs that implement a subset of the \texttt{struct\_ops} interface (\cref{tab:cca-hooks}), and (2) operators and service owners specify policies that define which connections the new CCA should be applied to. These policies specify both the locality (i.e., cross-region versus intra-region) and the scope (i.e., the services) of the connections. Given these two inputs, \sysname manages the lifecycle of the CCA's programs (i.e., installs them) and configures them (i.e., installs the appropriate configuration in their maps), which allows the CCA to be correctly applied to connections across our fleet.

\subsubsection{BPF-TCP Interface Evolution}
\label{sec:modular:hooks:evolution}
\vspace{\subsecvspace}

\Cref{tab:cca-hooks} shows the hookpoints that each of our CCAs implements; the rows for R1 and R2 show how we extended the interface to support more advanced CCAs. This highlights a core limitation: developers can introduce a new CCA only if they can realize it within the current interface.

\begin{table*}[t]
\centering
\scriptsize
\begin{tabular}{lllcccc}
\toprule
\textbf{Hook} & \textbf{Purpose} & \textbf{Release} & \textbf{BPF-DCTCP} & \textbf{\DTCP} & \textbf{\Glide} & \textbf{BPF-Cubic} \\
\midrule
\texttt{init}           & Initialize CCA state              & R0 & \checkmark & \checkmark & \checkmark & \checkmark \\
\texttt{ssthresh}       & Set slow-start threshold           & R0 & \checkmark & \checkmark & \checkmark & \checkmark \\
\texttt{set\_state}     & React to TCP state changes         & R0 & \checkmark & \checkmark & \checkmark & \checkmark \\
\texttt{undo\_cwnd}     & Undo spurious CWND reductions      & R0 & \checkmark & \checkmark & \checkmark & \checkmark \\
\texttt{cwnd\_event}    & React to CWND-related events       & R0 & \checkmark & \checkmark & \checkmark & \checkmark \\
\midrule
\texttt{cong\_avoid}    & Window-based congestion avoidance  & R0 & \checkmark & \checkmark &            & \checkmark \\
\texttt{in\_ack\_event} & Extract per-ACK congestion signals & R0 & \checkmark & \checkmark &            &            \\
\texttt{pkts\_acked}    & Per-RTT delivery-rate statistics   & R1 &            & \checkmark &            & \checkmark \\
\texttt{cong\_control}  & Full control over CWND and pacing  & R1 &            &            & \checkmark &            \\
\texttt{min\_tso\_segs} & Override TSO segment count         & R2 &            & \checkmark & \checkmark & \checkmark \\
\bottomrule
\end{tabular}
\caption{BPF-TCP hooks and their usage across four CCAs. The top five hooks (R0) are used by all four CCAs; the remaining five are used selectively. }
\label{tab:cca-hooks}
\end{table*}

\textbf{Limitations.} The set of CCAs that can be realized is restricted by the interface that \texttt{struct\_ops} exposes. First, the interface supports delay, ECN, and loss signals, but it does not provide richer in-band network telemetry (INT) signals. We note that for a subset of these INT signals, the interface may provide sufficient information to reconstruct them within one of its eBPF programs. Second, the BPF-TCP paradigm enables core CCA functionality and, more recently, cross-layer functionality. However, we found that certain performance-critical socket-level functionality falls outside the core CCA logic and was thus not configurable. While we exposed a subset of this functionality, much of it remains unexposed. As we show below, the community is open to evolving this interface when specific use cases justify it.

\textbf{R0: Core CCA Logic.}
The top five hooks provide control over the core CCA logic: initialization, state management, and basic congestion response. Our initial eBPF-based CCAs, BPF-Cubic and BPF-DCTCP, used these core hooks as well as \texttt{cong\_avoid} and \texttt{in\_ack\_event}. We were also able to extend Cubic into our transitional CCA by modifying one of its BPF programs to support ECN.

\textbf{R1: Efficient Per-RTT Signals and Control over Socket-Level Functionality.}
First, we initially designed \DTCP to infer per-RTT metrics from individual ACK events; however, this incurred significant overhead (compute to calculate per-RTT metrics and memory to maintain this state). We observed that the kernel already precomputes per-RTT metrics (e.g., \texttt{min\_rtt} and \texttt{max\_send\_rate}), so we modified the hook signature to expose them.
Second, we noticed that the kernel overrode certain decisions without the CCA's knowledge, e.g., forcing a connection into the CWR state on a single ECN mark or imposing send-buffer growth policies. We extended the interface with a new hookpoint that provides programmatic control over this internal socket-level functionality.

\textbf{R2: Control over Cross-Layer Functionality.} Finally, we found that we needed control over certain properties not traditionally considered CCA functionality, e.g., \texttt{tcp\_min\_tso\_segs}. Our last change extends the BPF-TCP interface to provide these controls.

\subsection{Cross-Program State Sharing}
\label{sec:modular:sharing}
\vspace{\subsecvspace}
To ensure consistent behavior across the programs within a BPF-TCP \texttt{struct\_ops} bundle, and to enforce correct policy across our CCA programs and other eBPF programs, we introduced a broad set of state-sharing mechanisms.

\subsubsection{Coordination Across Our Fleet}
\label{sec:modular:sharing:pipeline}
\vspace{\subsecvspace}

\begin{figure}[t]
    \centering
    \includegraphics[width=0.85\linewidth]{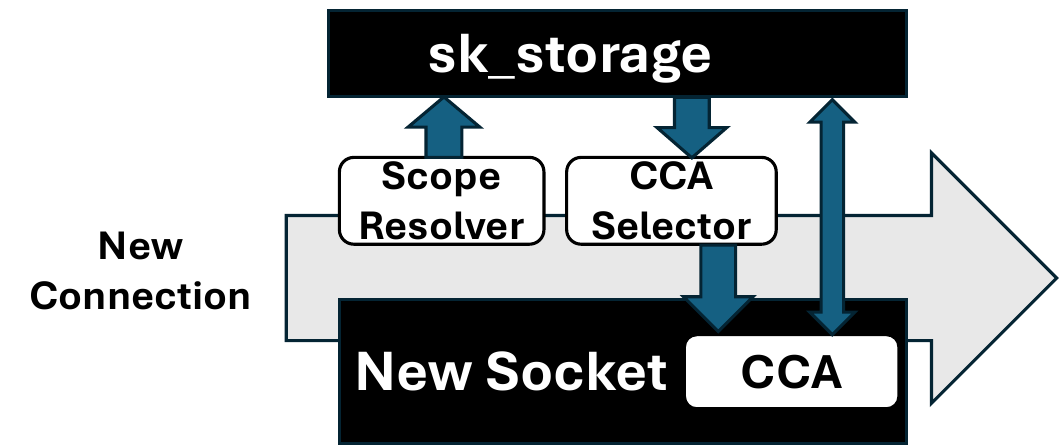}
    \caption{\sysname pipeline for configuring the CCA for a new connection and its socket. }
    \label{fig:newconnection}
\end{figure}

\sysname dynamically configures and sets the CCA for each connection using a pipeline (\cref{fig:newconnection}) of eBPF programs developed by several teams. Reusing these programs ensures consistent policy across teams.

\begin{itemize}[leftmargin=*, nosep]
    \item First, NetEdit's scope resolver (a sockops program) performs a longest-prefix match on the remote address to identify the connection's locality (i.e., intra-region, cross-region, or unknown) and stores the result in an \texttt{sk\_storage} map.
    \item Next, NetEdit's CCA selector (a sockops program) reads the connection's scope from the \texttt{sk\_storage}  map and then calls \texttt{bpf\_setsockopt} to set the connection's CCA, assigning the appropriate \texttt{struct\_ops} to the connection's socket.
    \item The CCA itself (a \texttt{struct\_ops} program) reads the scope from the same \texttt{sk\_storage} entry to retrieve per-locality parameters.\footnote{\DTCP additionally consumes one-way queuing delay (OWD) measurements written per packet by a separate program owned by a different team.}
    \item Finally, as services migrate, \sysname must ensure that the policies stored in the policy configuration maps\footnote{The configuration map is keyed by source IP, destination region, and destination data center, allowing different services on the same host to use different CCAs with different parameters.} are updated to reflect the current set of services on a host.
    To do this, the scope resolver repopulates the scope maps, and the CCA selector dynamically re-evaluates each new connection based on the updated scope maps.
\end{itemize}

\begin{table}[h]
\centering
\scriptsize
\begin{tabular}{llll}
\toprule
\textbf{State} & \textbf{Producer (attach, rate)} & \textbf{Consumer (attach, rate)} & \textbf{Store} \\
\midrule
Scope   & Resolver (sockops, once)   & Selector, CCA (per socket) & \texttt{sk\_storage} \\
CCA name & Selector (sockops, once)  & Kernel TCP (per socket) & socket \\
Config  & Control plane (minutes)    & Selector, CCA (per socket) & global map \\
\bottomrule
\end{tabular}
\caption{State sharing within the \sysname CCA configuration pipeline.}
\label{tab:pipeline}
\end{table}

\textbf{State Within a CCA \texttt{struct\_ops}.} We use the \texttt{sk\_storage} map type because it is indexed by socket pointer; since CCA \texttt{struct\_ops} programs have the connection's socket pointer, CCA-related lookups take O(1) time. In a microbenchmark on a production host, we found that an \texttt{sk\_storage} lookup costs ${\sim}$25\,ns, versus ${\sim}$1\,$\mu$s for an LPM scope lookup. Additionally, \texttt{sk\_storage} maps automatically garbage-collect entries when sockets are destroyed (i.e., when connections close), which eliminates the complex garbage collection that other map types require.

\textbf{State Across a CCA \texttt{struct\_ops} and Other eBPF Programs.}
Shared state is the only way to compose or coordinate CCA BPF programs with other BPF programs. Unfortunately, due to eBPF design constraints, \texttt{sk\_storage} maps cannot be used by kernel modules or userspace, so we must build shared storage from other map types to coordinate across programs. In particular, NetEdit and its programs populate and manage the policy configuration maps, and \sysname's CCA programs read and use this configuration. These policy configuration maps are an example of shared state that cannot be stored in \texttt{sk\_storage}.

\subsubsection{CCA-Specific State-Sharing Edge Cases}
\label{sec:modular:sharing:constraints}
\vspace{\subsecvspace}
\Cref{sec:modular:sharing:pipeline} cleanly assigns different types of eBPF maps to different roles within \sysname; however, at scale we encountered a set of edge cases that forced us to make several compromises.

\textbf{Splitting Per-Connection State.}
In our designs, the CCA-private (i.e., per-connection) state was much larger than the kernel's existing data structure can hold (i.e., the 104 bytes provided by \texttt{icsk\_ca\_priv}). Consequently, we split this state across two data structures: (1) the internal CCA structure and (2) \texttt{sk\_storage}. This split lets us support larger (and more complex) per-connection logic, but at the cost of one additional lookup per read, because we must look up the \texttt{sk\_storage} entry in addition to the internal structure.

\textbf{LPM Fallback.}
\sysname relies on one of NetEdit's programs, the scope resolver, to populate certain maps. If these maps are unavailable, \sysname transparently performs an LPM lookup and caches the result in the appropriate maps for future connections. This ensures that a CCA's correctness is independent of the scope resolver's availability. We note that we designed NetEdit and its programs to be highly available so that this LPM fallback is infrequent, because it is expensive and may degrade application performance.

\subsection{\sysname Use Cases}
\label{sec:modular:usecases}
To satisfy our unique deployment constraints, we generally evolve and adapt existing algorithms (e.g., BBR, Swift, DCTCP, and Cubic). In this section, we discuss two CCAs that we recently designed and show how their designs justify extending BPF-TCP beyond the initial interface (i.e., R0).

\subsubsection{DCNDelay-CCA: Delay-Based Control for Heterogeneous Data Centers}
\label{sec:modular:d2tcp}
\vspace{\subsecvspace}

We introduced a new delay-based CCA, \DTCP, to replace DCTCP~\cite{dctcp}. Two key factors motivated this move. First, correctly configuring ECN across our heterogeneous fleet was challenging, and the fleet's continual evolution made configuring ECN even more time-consuming. Second, while we experienced significant congestion at the host, many host NICs do not mark ECN bits, leaving DCTCP blind to a significant source of contention.
In contrast, delay provides a unified signal of queuing regardless of where it occurs (network switches, NIC buffers, or the host PCI bus) and requires minimal configuration.

\textbf{Algorithm.}
At a high level, \DTCP maintains low queue occupancy by computing a target congestion window from the observed delay and a configured minimum RTT. When the network is free of queuing, the target window grows linearly; when queuing is observed, the algorithm reduces the target window. We deliberately chose a simpler design than other delay-based approaches (e.g., BBR) because we wanted to avoid designing additional heuristics to tailor a complex design to our diverse hardware ecosystem. \DTCP also uses ECN as a fallback to ensure that it is never more aggressive than DCTCP. Further details of the algorithm appear in \cref{appendix:d2tcp:algorithm}.

\textbf{Hookpoints Used.}
\DTCP's core algorithm uses three hookpoints (\cref{tab:cca-hooks}): \texttt{in\_ack\_event} to compute per-ACK ECN alpha and delay updates; \texttt{pkts\_acked} to capture per-RTT statistics and process loss; and \texttt{min\_tso\_segs} to dynamically configure the TSO segment count so that the CCA is not bottlenecked by NIC behavior.

\subsubsection{XRegion-CCA: ECN-Driven Control for Cross-Region WANs}
\label{sec:modular:xregion}
\vspace{\subsecvspace}

We replaced Cubic with \Glide as the CCA for our cross-region traffic. We opted for ECN rather than delay as the cross-region congestion signal because our experience showed that cross-region delays change frequently due to our traffic-engineering system~\cite{meta:ebb}, which regularly rebalances virtual paths across physical links. Because this rebalancing is invisible to end hosts, an observed change in delay could be due either to congestion or to rebalancing. We therefore opted for an ECN-based CCA, which we designed to interoperate with \DTCP's shallow buffer occupancy and with flows from other data centers that operate at different RTTs.

\textbf{Algorithm Design.}
\Glide builds on Swift's~\cite{google:swift} control loop and adapts it to the ECN signal. Recall that Swift is a delay-based CCA that compares a per-RTT delay measurement against a target. When delay is below the target, Swift grows the window additively; when delay is above the target, Swift reduces the window multiplicatively, in proportion to the overshoot. \Glide replaces delay with the \emph{fraction of ECN-marked ACKs} and compares it against a target fraction. Further details of the algorithm appear in \cref{appendix:xregion:algorithm}.

\textbf{Hookpoints Used.}
\Glide uses the \texttt{cong\_control} hookpoint to control how the kernel processes and reacts to ECN marks.\footnote{Recall that the kernel forces a connection into the CWR state on a single ECN mark; \Glide must override this behavior to function correctly.} Additionally, for efficiency, \Glide uses the \texttt{min\_tso\_segs} hookpoint to calculate and maintain efficient segment batching. Finally, \Glide implements the \texttt{init} hookpoint to minimize convergence time by setting the initial congestion window (we use a default of 100 MSS).

\section{\sysname Orchestrator}
\label{sec:ecosystem}
\vspace{\subsecvspace}

This section describes the challenges of, and our approach to, deploying, updating, and operating CCAs at fleet scale. Our approach inherits certain properties from our eBPF orchestration platform, NetEdit: specifically, disruption-free deployment (\cref{sec:ecosystem:deployment}), retroactive application, and safe rollback. We introduce a transitional CCA that NetEdit can safely roll back to when a CCA deployment fails (\cref{sec:ecosystem:transitionalcca}). We also extend NetEdit to support coordinated CCA orchestration (\cref{sec:ecosystem:orchestration}).

\subsection{Guarantees Provided by NetEdit}
\label{sec:ecosystem:deployment}
\vspace{\subsecvspace}

\begin{figure}[t]
    \centering
    \begin{minipage}{0.48\linewidth}
        \centering
        \includegraphics[width=\linewidth]{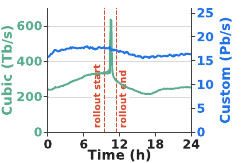}
        \vspace{0.3em}
        \textbf{(a)} Without object pinning.
        \label{fig:iter:norollout}
    \end{minipage}
~
    \begin{minipage}{0.48\linewidth}
        \centering
        \includegraphics[width=\linewidth]{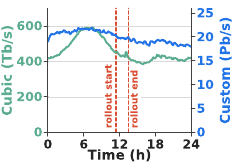}
        \vspace{0.3em}
        \textbf{(b)} With object pinning.
        \label{fig:after:rollout}
    \end{minipage}

    \caption{CCA deployment in a region: (a) without disruption-free rollout, a visible Cubic spike appears during upgrades; (b) with disruption-free rollout, the spike is eliminated.}
    \label{fig:free:rollout}
\end{figure}

\textbf{Disruption-Free Deployment.} At Meta, we run an eBPF orchestrator, NetEdit~\cite{meta:netedit}, which deploys new versions of our eBPF programs (including our eBPF-based CCAs) to production every week. During an upgrade, NetEdit is briefly unavailable on each host (typically ${\sim}$30 seconds), and any new connection established during this window falls back to the kernel's default CCA (i.e., Cubic). Although this window is short, at scale a large number of connections end up using this default, which performs suboptimally relative to our tailored CCAs.
To eliminate this disruption, we apply NetEdit's BPF pinning mechanism~\cite{meta:netedit}, which decouples the lifecycles of eBPF objects (e.g., maps and programs) from the control-plane process that manages them. This decoupling ensures that when an update restarts the control-plane process, the eBPF objects continue running unaffected and remain configured to use the appropriate optimized CCAs.
\Cref{fig:free:rollout} highlights the benefit of building \sysname on NetEdit and leveraging its object management. Without object pinning (\cref{fig:free:rollout}(a)), a large amount of new traffic uses Cubic during the upgrade, producing a visible spike in Cubic traffic; with NetEdit's object pinning (\cref{fig:free:rollout}(b)), the spike is eliminated.

\textbf{Retroactive Application.} After a new CCA is deployed, we must migrate all existing connections that use the old version to the new CCA. We do this using NetEdit's TCP iterator (\texttt{bpf\_iter})~\cite{meta:netedit}, which enumerates all active connections and applies the new CCA's logic to each one. As described in NetEdit, this mechanism assumes that CCA logic is \emph{idempotent}: reapplying initialization or policy logic does not violate correctness.\footnote{While we expect changes to be nearly instantaneous, certain state variables do not update until further packets are processed; \cref{sec:evaluation:deployment} quantifies the transient.} The TCP iterator provides an added benefit: because it is applied to all connections, it also updates connections whose shared inputs (e.g., scope, locality, or CCA configuration) have gone stale (\cref{sec:modular:sharing:constraints}).

\subsection{Introducing a Transitional CCA for Safe Rollback}
\label{sec:ecosystem:transitionalcca}
\vspace{\subsecvspace}
When a newly rolled-out CCA is buggy, we want to roll back to a trusted CCA, i.e., a transitional CCA, that can be safely applied to traffic using any other CCA. Designing a transitional CCA is challenging because certain CCAs rely on protocol features negotiated in advance, so rollback must revert to a CCA that is compatible with these prenegotiated features (e.g., once negotiated, ECN persists even after the CCA is changed, so the transitional CCA must support ECN). Unfortunately, Cubic (our default CCA) is not designed for ECN and treats every ECN mark as it would a loss. We therefore extended Cubic into our transitional CCA (BPF-Cubic) by modifying its code to handle ECN signals correctly.
In practice, rollback follows the same three broad steps as NetEdit's rollout procedure: (1) the transitional CCA is loaded as the new active CCA, (2) the TCP iterator retroactively applies it to all existing connections, and (3) the faulty CCA is unregistered.

\subsection{Supporting CCA-Centric Orchestration}
\label{sec:ecosystem:orchestration}
\vspace{\subsecvspace}
While NetEdit is designed to roll out eBPF programs to hosts within our fleet, this approach fails for CCA rollouts in which both ends of a connection must use the same CCA.\footnote{Certain CCAs (e.g., ECN-based ones) require \emph{both} endpoints to be upgraded simultaneously; e.g., ECN must be negotiated during the TCP handshake.} We therefore extended NetEdit to support coordinated rollouts.

For intra-region CCAs, rollout is straightforward and uses existing NetEdit primitives, which upgrade all servers at a specified granularity together. We upgrade all servers within a region at once, ensuring that the source and destination of every intra-region connection use the same CCA.

For cross-region CCAs, we added a mechanism to NetEdit that supports \emph{region-pair} rollouts. We roll out to pairs of regions and update the scope resolver so that cross-region connections between upgraded region pairs use the new CCA.
\begin{figure}[t]
    \centering
    \includegraphics[width=0.95\linewidth]{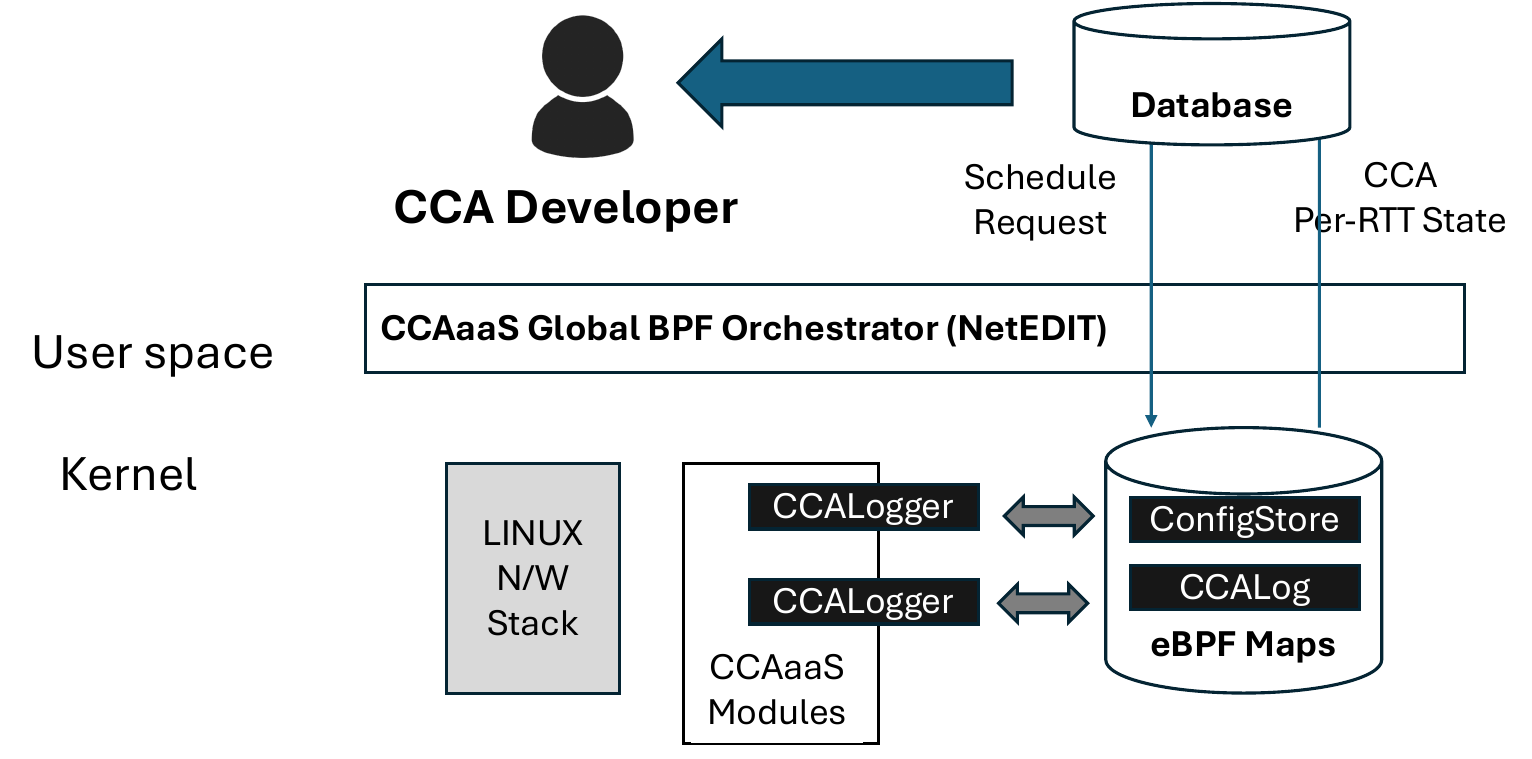}
    \caption{Workflow of \CCALogger in production. }
    \label{fig:ccalogger:workflow}
\end{figure}

\section{\CCALogger}
\label{sec:ccalogger}

We designed \CCALogger to provide fine-grained visibility into eBPF-based CCAs by tracing CCA-specific state and signals over the lifetime of a dynamically selected connection. To scale \CCALogger to production environments, our design carefully manages state and control logic across many concurrent connections while working within a strict performance budget and the kernel's limited support for synchronization primitives. \CCALogger addresses these challenges through three generalizable principles:

\begin{itemize}[leftmargin=*, nosep]
    \item \textit{Align observability with control timescales.}
Analyzing CCA internals or evaluating logging conditions at per-packet granularity is prohibitively expensive. Our design builds on the insight that important CCA state evolves at RTT granularity; thus, as long as we log once per RTT, we faithfully capture the control dynamics. Consequently, \CCALogger never logs state at a finer granularity than once per RTT, which enables it to scale.
    \item \textit{Couple observability with modular CCA components.}
Given per-RTT logging, the next challenge is deciding \emph{when} within an RTT to log in order to capture the most accurate snapshot. Because our CCA framework is modularized to invoke eBPF programs at well-defined state transitions and signal changes, the natural observation point is the existing CCA function that performs the per-RTT logic. Consequently, \CCALogger integrates into the specific BPF programs that perform per-RTT analysis and control.
    \item \textit{Decouple policy from mechanism.}
The set of connections to monitor varies with workload and context; e.g., our CCA developers often wish to analyze connections based on service identity, endpoints, or previously observed behavior. Unfortunately, supporting this flexibility within the kernel would incur significant overhead, because it requires evaluating predicates (i.e., comparisons and conditional branches) on the datapath. Instead, \CCALogger performs all policy logic in user space and uses a simple selection mechanism in the kernel.
\end{itemize}

\subsection{Architecture}
\label{sec:architecture}
\CCALogger (\cref{fig:ccalogger:workflow}) realizes these principles by confining policy analysis and decisions to user space while keeping the datapath minimal, predictable, and efficient. This control and data plane split mirrors the philosophy of \sysname and more generally eBPF orchestration at Meta (i.e., NetEdit).

\textbf{Control Plane.}
The control plane decides \emph{what} to observe. Developers specify the connection, or class of connections, to monitor using a CLI or web interface. This specification is distributed across the fleet via our configuration management system. On each host, a user-space agent translates the specification into compact selection state and updates the corresponding in-kernel data structures (e.g., the configuration map). The agent periodically retrieves the logged CCA state, stores it in a local database, and exposes it to developers through the CLI or web interface.

\textbf{Data Plane.}
The data plane decides \emph{when} and \emph{how} to record state. It comprises a small set of functions inserted into the BPF-TCP programs that are triggered at RTT boundaries and that have access to stable CCA control state. Specifically, we instrument \texttt{init} to record static connection metadata and \texttt{pkts\_acked} to capture per-RTT CCA state.\footnote{Although \texttt{pkts\_acked} may be invoked multiple times per RTT, logging is coded so that state is written exactly once per RTT.}

Each function performs a constant-time ownership check and, if the check succeeds, exports CCA state to a log. The datapath relies on two shared data structures: a configuration map used for the ownership check and a log used to record state.
The ownership check is how \CCALogger minimizes datapath overhead: it reduces the computationally expensive selection logic to a single 64-bit comparison.\footnote{A naive design would encode selection as predicates over internal CCA state or connection attributes (e.g., the 5-tuple or process identifiers), requiring continuous evaluation and significantly increasing datapath complexity.} Specifically, the function compares the socket cookie of the connection it is processing against a value stored in the configuration map. Socket cookies are compact, stable identifiers that uniquely identify a socket for its lifetime. We chose socket cookies over alternatives, e.g., socket pointer addresses, because they are explicitly exposed to user space (i.e., our control plane).

\subsection{\CCALogger Use Cases}
\vspace{\subsecvspace}
We next briefly describe our two most common uses of \CCALogger.

\textbf{Development and Debugging.}
A key challenge we faced while developing our BPF-based CCAs was validating whether a specific implementation operates correctly. \CCALogger exposes a CCA's internal structures and state values and how they change over time, which also lets us analyze state transitions. For example, during the development of \Glide, \CCALogger revealed that the kernel's default state machine immediately transitioned a sender from the OPEN to the CWR state upon receiving a single ECN mark. This revealed the need to override the kernel's state machine by implementing the \texttt{cong\_control} hook (\cref{sec:modular:hooks:evolution}).

\textbf{Operational Validation.}
We also use \CCALogger to monitor our CCAs in production, ensuring that connections use the appropriate CCA and exhibit the intended behavior. For instance, while deploying \Glide, we used \CCALogger to monitor compliance and, specifically, to confirm that over 90\% of active connections ran \Glide rather than falling back to the kernel default.

\section{Operational Lessons}
\label{sec:operational}
\vspace{\subsecvspace}
Recent work has characterized lessons from managing and scaling eBPF deployments~\cite{BPFComposer,meta:netedit,bytedance:netkit}.
\footnote{Portability refers to running the same program across different kernel versions, which is challenging because eBPF support is missing or inconsistent across versions.} In light of this work, we focus on CCA-specific challenges: specifically, lessons from modular CCA design (\cref{sec:operational:modularity}), from managing eBPF-based CCAs (\cref{sec:operational:management}), and from evolving eBPF to support CCA development (\cref{sec:operational:ebpf}).

\subsection{Lessons for Designing Modular CCAs}
\label{sec:modular:lessons}
\label{sec:operational:modularity}
\vspace{\subsecvspace}
We have identified important lessons from experience that research on modular CCAs~\cite{mtcp,ccp,tas,CTP:ToN07,mtp:arxiv25,tran2019beyond,etran} often does not address.

\textbf{Prefer specialized interfaces over generalized ones.}
A class of research frameworks argues for generalized event-based models; however, in our experience, abstract interfaces force functions to recompute statistics and infer CCA properties, incurring significant overhead (in both compute and storage). Naturally, specialized interfaces introduce some ossification; however, we found the BPF-TCP interface amenable to evolution.

\textbf{Provide control over socket-level functionality, kernel configurations, and cross-layer network stack functions.}
Research frameworks~\cite{ccp,mtp:arxiv25,etran} focus on the core CCA control loop, overlooking kernel behaviors below the CCA (e.g., TSO batching and buffer growth heuristics); however, we found that these layers silently interact with CCA decisions and in certain cases override CCA behavior. Thus, frameworks should provide control over these elements.

\textbf{Evolve CCAs incrementally.}
In production, new CCAs must be deployed without impacting the performance or availability of existing applications. This is challenging and requires that new CCAs be incrementally deployable and interoperable. Fundamentally, clean-slate approaches, as often proposed in research, are disruptive and require changes to existing applications and infrastructure. A natural path is to take existing CCAs and gradually evolve them to support new requirements and to adapt to lessons learned from production deployments.

\subsection{Lessons for Managing CCAs at Scale}
\label{sec:operational:transport}
\label{sec:operational:management}
\vspace{\subsecvspace}
Unlike the previous lessons, which focus on designing a modular CCA framework, this set of lessons focuses on
the operational challenges we faced while continuously rolling out CCAs without impacting applications.

\textbf{Explicitly override the kernel default with your most performant CCA.} Our initial approach used the kernel's default CCA, i.e., Cubic, for connections with unknown scope or locality. However, as we expanded applications into public clouds for inference, these connections had unknown scope and were therefore configured to use Cubic, a CCA known to be suboptimal for WAN traffic. We updated NetEdit to include locality and scope configuration for these cloud infrastructures. More generally, we now select the default based on our most performant CCA rather than the kernel default.

\textbf{Choose your rollout granularity for both problem attribution and blast radius.}
During our initial rollout of \Glide, we upgraded a subset of connections to use \Glide, which meant that connections from the same service used different CCAs. Unfortunately, our network telemetry aggregates metrics (e.g., retransmission rates and ToR buffer occupancy), so we could not debug \Glide: we could not attribute aggregate behavior to \Glide or to other CCAs.
We enhanced NetEdit's rollouts to support region-pair granularity, so that all traffic to a given region uses the same CCA. This allowed us to attribute changes in metrics to the CCA change.

\textbf{Plan for network state and bugs.}
Our initial deployment of \Glide when it negotiated ECN, the WAN routers incorrectly interpreted it as  DSCP. When we rolled back to BPF-Cubic, we ran into further problems because \Glide had already negotiated ECN. We learned to explicitly account for the network by testing any new network features a CCA uses before deploying it. Similarly, we designed our transitional CCA to support prenegotiated network state.

\subsection{Lessons from Evolving eBPF to Support TCP}
\label{sec:operational:platform}
\label{sec:operational:ebpf}
\vspace{\subsecvspace}

We designed NetEdit for sockops and TC programs; however, extending it to support \texttt{struct\_ops} CCAs revealed several issues that, we believe, modern orchestration frameworks must robustly account for as eBPF expands to more hookpoints.

\textbf{Verifier-approved programs can still take down the kernel.} In using BPF-TCP, we identified latent kernel bugs that disrupted production traffic. Our programs triggered these bugs despite passing the eBPF verifier's checks and being considered safe. The bugs ranged from recursion bugs and reference-count corruption to return-value bugs. Most interestingly, these bugs were often nondeterministic, both in when they occurred and in the symptoms they produced. We patched these bugs~\cite{BPFComposer,patch1,patch3,patch4}.

\textbf{Validate that hookpoint semantics hold for new hookpoints.} We designed NetEdit to explicitly tackle heterogeneity in kernel semantics by decoupling eBPF behavior from kernel heuristics and introducing an interface that lets NetEdit explicitly manage the eBPF lifecycle. However, our design assumed that hookpoints share certain invariant behaviors that we could design around. This held until we extended NetEdit to \texttt{struct\_ops}: we assumed that eBPF programs are managed by a single userspace entity, but \texttt{struct\_ops} objects are managed by both userspace and the kernel's CCA subsystem. Consequently, terminating the userspace process does not remove \sysname's eBPF objects, and a redeployed userspace process fails to load new eBPF objects; when this happened, \sysname could not configure the CCA for connections on that host.\footnote{Binding each \texttt{struct\_ops} object to a \texttt{bpf\_link} owned by NetEdit, so that the CCA is reclaimed when the link's owner exits, reduced registration errors from ${\sim}$4K hosts per week to single digits.} We therefore bound each \texttt{struct\_ops} object to a \texttt{bpf\_link}, which allows it to be managed entirely by one entity. We are now working to identify other such invariants, and we welcome tools that automatically characterize these invariants and check them as new hookpoints are adopted.

\textbf{Introduce versioning semantics for shared state.}
When the kernel loads a program, it validates only that the map's key and value sizes match the program's definitions. When we rolled back a program, it wrote well-formed but version-incompatible data, which caused our CCA to behave incorrectly for ${\sim}$25\% of our traffic and degraded performance. We now version all of our maps to avoid such issues.

\subsection{Upstream Contributions}
\label{sec:operational:patches}
\vspace{\subsecvspace}
We upstreamed several kernel patches as part of this project: (1) \texttt{struct\_ops} link semantics for correct CCA lifecycle management~\cite{patch1}, which allow eBPF orchestrators to automatically clean up \texttt{struct\_ops} objects when they exit; and (2) a patch that enables live, zero-downtime connection migration between CCAs~\cite{patch3}.

\begin{table}[t]
\centering
\scriptsize
\begin{tabular}{lrrrrc}
\toprule
\textbf{CCA} & \textbf{Core} & \textbf{Headers} & \textbf{CCALogger} & \textbf{Total} & \textbf{re-used lines} \\
\midrule
BPF-DCTCP    & 209 & 50  & 130 & 389 & 73 \\
DCNDelay-CCA & 472 & 91  & 141 & 704 &  132  \\
XRegion-CCA  & 440 & 350 & 127 & 917 & 253  \\
BPF-Cubic    & 182 & 46  & 132 & 360 & 0  \\
\bottomrule
\end{tabular}
\caption{Lines of code per CCA and their provenance.}
\label{tab:cca-loc}
\end{table}

\section{Implementation}
\label{sec:implementation}
in this section, we briefly describe the implementation of our CCAs: our CCAs are in a combination of C file and header files. Table~\ref{tab:cca-loc}, illustrates the breakdown of the different CCAs as well as the re-use across CCAs due to the modularity that Struct\_ops provides. For example, our implementation of \DTCP extends on our existing BPF-DCTCP with two additional hooks and ${\sim}$340 lines of code. More importantly, we note that a non-trivial fraction of our newer CCAs reuse code from our earlier CCAs enabling the rapid development cycles highlighted in Table~\ref{tab:cca_timeline}.

\section{Evaluation}
\label{sec:evaluation}
\vspace{\subsecvspace}
Our evaluation of \sysname is motivated by the following questions. What are the overheads (both computational and performance) of implementing a CCA in eBPF (\cref{sec:evaluation:cost})? What are the benefits of the new CCAs enabled by \sysname (\cref{sec:evaluation:validation})? Do CCAs built with our modular framework achieve production-quality performance (\cref{sec:evaluation:validation})? What are the benefits of the management primitives introduced by \sysname (\cref{sec:evaluation:deployment})?

\textbf{Evaluation Setup.} We collected metrics from running \sysname across environments with varying traffic types (streaming and RPC), flow counts, sender counts, NIC speeds (50G and 100G), NIC vendors, and scopes (intra-rack, intra-data center, intra-region, and cross-region). For a subset of experiments (\cref{sec:evaluation:cost}), we selected a rack on which to explore different CCAs without disrupting production traffic. We evaluated the overhead of eBPF-based CCAs using 49 production hosts running ${\sim}$1{,}100 TCP connections. In each experiment, all hosts use the same CCA.

\subsection{Analysis of eBPF-based Modularity}
\label{sec:evaluation:cost}
\vspace{\subsecvspace}

\begin{figure}[h]
    \centering
    \includegraphics[width=\columnwidth]{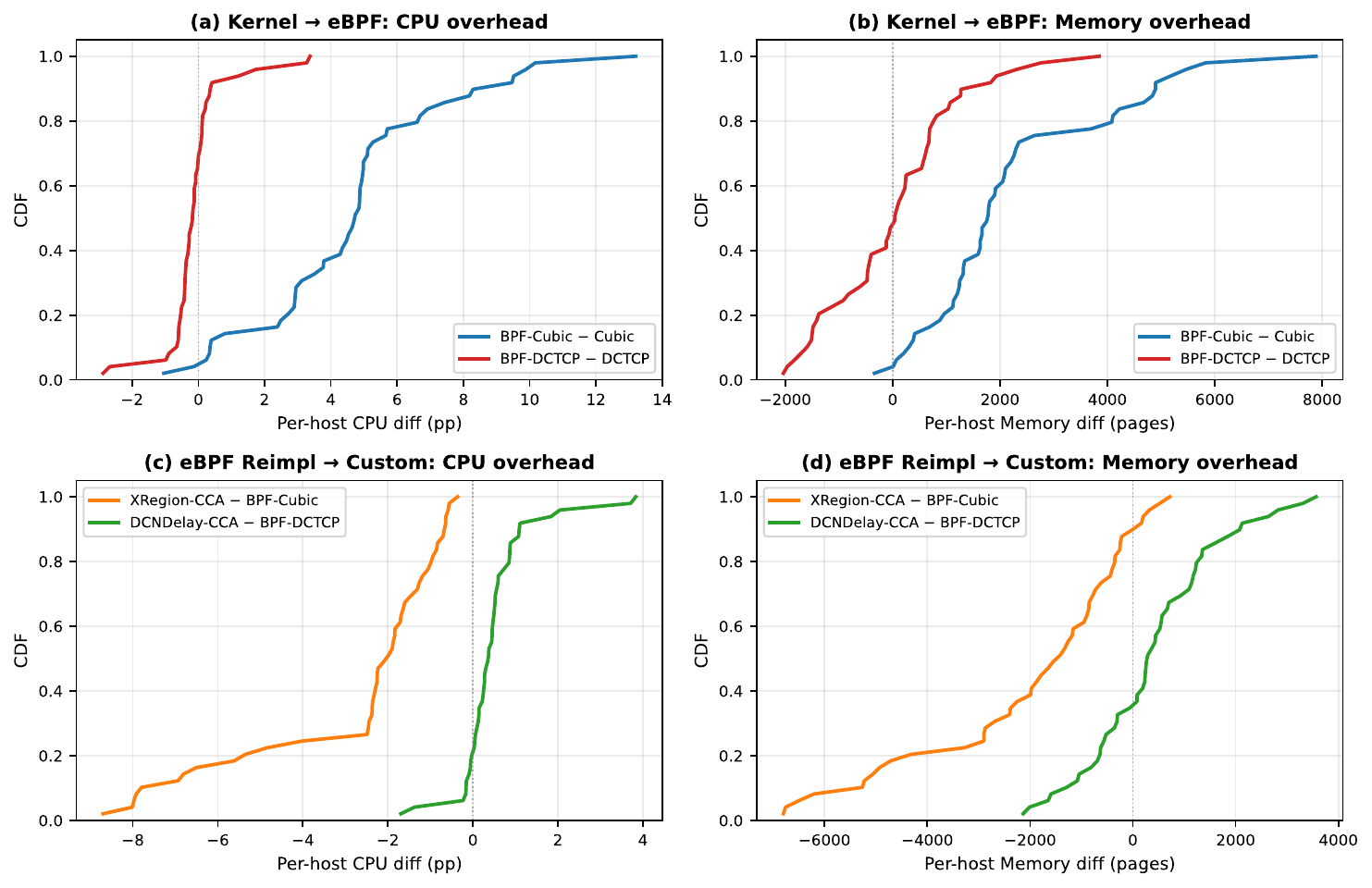}
    \caption{Paired per-host CDFs of CPU and memory differences. We exclude application-level metrics (retransmissions, bandwidth, and P99 latency) because they are virtually identical. CPU graphs are percentage points (PP).}
    \label{fig:paired:cdfs}
\end{figure}

In \cref{fig:paired:cdfs}(a) and (b), we compare native kernel and eBPF-based variants of DCTCP and Cubic. We observe negligible differences, which shows that implementing a CCA with the BPF-TCP interface yields no meaningful performance difference compared with implementing it directly in the kernel.
However, BPF-Cubic introduces ${\sim}$5~pp CPU overhead, attributable to the eBPF JIT dispatch cost being a larger fraction of Cubic's lightweight arithmetic.

\subsection{ Analysis of \sysname-Enabled CCAs}
\label{sec:evaluation:validation}
\vspace{\subsecvspace}
We present deployment results for our two most recent CCAs to validate the framework's efficiency and utility.
For context, we replaced DCTCP with \DTCP as the default for over 94\% of our intra-region traffic, and we similarly replaced Cubic with \Glide for 94.7\% of all cross-region traffic.

\textbf{Relative Overheads of New CCAs.}
We analyze the impact of our extensions to the BPF-TCP interface by comparing the overheads of the new CCAs against eBPF implementations of traditional CCAs (\cref{fig:paired:cdfs}(c) and (d)). We observe that \DTCP adds only 0.4~pp CPU over BPF-DCTCP, while \Glide actually reduces CPU by 2.0~pp relative to BPF-Cubic.

\begin{table}[h!]
\centering
\scriptsize
\begin{tabular}{l|cc|cc}
\toprule
\textbf{CCA}
& \multicolumn{2}{c|}{\textbf{CPU Overhead (\%)}}
& \multicolumn{2}{c}{\textbf{Memory (bytes)}} \\
& \textbf{p95} & \textbf{p99} & \textbf{p95} & \textbf{p99} \\
\midrule
\DTCP  & 3  & 26  & 45,056  & 269,864 \\
\Glide & $<$1 & 1  & 45,056  & 185,964 \\
\bottomrule
\end{tabular}
\caption{Resource overheads (p95 and p99) of recent CCAs.}
\label{tab:cca_overhead_percentiles_grouped}
\end{table}

\textbf{Overheads of New CCAs.}
Next, in \cref{tab:cca_overhead_percentiles_grouped}, we analyze these CCAs' fleet-wide resource footprint. We observe that both CCAs impose minimal CPU and memory overhead. Because \sysname maintains per-connection state,\footnote{Each \texttt{struct\_ops} CCA has a fixed overhead of ${\sim}$12KB (the BPF program and counter map) plus 104 bytes per active connection (\cref{sec:modular:sharing:constraints}).} the memory variation across our fleet can be attributed to variation in the number of concurrent connections per host.

\begin{table}[h!]
\centering
\scriptsize
\begin{tabular}{lll}
\toprule
\textbf{CCA} & \textbf{Metric} & \textbf{Improvement} \\
\midrule
\multirow{4}{*}{\DTCP} & Retransmissions (Ads/Feed/Async) & 30--40\% reduction \\
& Intra-region congestion window & $>$50\% reduction \\
& ToR packet discards & Significant reduction \\
& Host CPU utilization & ${\sim}$5\% reduction \\
\midrule
\multirow{4}{*}{\Glide} & Congestion discards (fleet-wide) & $>$90\% reduction \\
& WRED packet drops (single region) & ${\sim}$40$\times$ reduction \\
& Cross-region service P99 latency & 4--20$\times$ improvement \\
& GenAI checkpoint P99 latency & ${\sim}$40$\times$ (20s $\rightarrow$ 0.5s) \\
\bottomrule
\end{tabular}
\caption{Production benefits of deploying \DTCP and \Glide.}
\label{tab:production_validation}
\end{table}

\textbf{Production Benefits of New CCAs.}
\Cref{tab:production_validation} summarizes the production improvements we observed after deploying \DTCP and \Glide.

For intra-region traffic, \DTCP significantly improved network metrics (retransmissions and packet discards), and these benefits translated into host-level benefits (i.e., lower CPU utilization).\footnote{The ${\sim}$5\% reduction in host CPU utilization, driven by larger effective packet sizes and fewer retransmissions, shows that the eBPF abstraction does not impose meaningful overhead.} Additionally, \DTCP's smaller congestion window reduced in-network queuing, creating headroom from which cross-region flows benefit (\cref{appendix:coexistence}).

For cross-region traffic, we also observed significant improvements in network metrics that translated into application-level improvements. Specifically, for the storage service that AI workloads use for checkpoint data, model weights, and configuration: \Glide significantly improved P99 latency over Cubic, especially for larger objects. For latency-sensitive workloads, we also observed that \Glide reduced service latency by up to 20$\times$.

\subsection{Analysis of Retroactive Application, and Rollback}
\label{sec:evaluation:deployment}
\vspace{\subsecvspace}

\textbf{Retroactive Application.}
During the initial \DTCP rollout, we used TCP iterator to migrate over 6 million existing sockets across one region's data centers. During this migration, we observed that per-host migration latency averaged approximately 4ms, with a P99 of 125--175ms. More importantly, we observed that TCP iterator produced no measurable impact on host-level metrics. However, we noticed that  migrated connections experience a temporary throughput decrease of 13\%  for 20--30ms while iterator modifies state variables which can then only be used by subsequent packets.

\textbf{Rollback.}
Rollback generally takes ${\sim}$3 minutes end-to-end. During the \Glide rollback, we reverted five regions to BPF-Cubic one region at a time rather than all at once, to prevent side effects from a sudden fleet-wide CCA change; each region completed within 5--10 minutes, covering all servers in its data centers, and all five regions were reverted within one day.

\section{Related Work}
\label{sec:related}
\vspace{\subsecvspace}

The work most closely related to \sysname falls into four main categories:

\textbf{Modular and Programmable CCA Frameworks.} Existing approaches to modular CCAs move logic to userspace~\cite{ccp,mtcp,tas} to enable rapid prototyping; instead, we use eBPF to enable rapid prototyping within the kernel, which allows us to reuse existing CCA mechanisms. Other work~\cite{etran,tran2019beyond} uses eBPF to implement CCAs; however, because these approaches use independent programs and focus on core CCA logic, they do not provide the level of management or customization that our CCAs require.

\textbf{Managing CCA Deployment at Scale.}
Hyperscalers have described their experiences deploying individual CCAs~\cite{metadctcp:nsdi24,google:bolt,google:swift,bytedance:barre,dcqcn}; however, these works generally focus on specific CCAs and their evaluation. Instead, we focus on the ecosystem of tools and frameworks for managing a large and evolving set of CCAs. Additionally, our insights focus on eBPF-based CCAs, whereas others have explored evolvability in the context of userspace stacks (e.g., Google's Snap~\cite{snap} and Alibaba's Luna~\cite{luna}).

\textbf{eBPF Management and Orchestration.}
Existing eBPF management systems are general-purpose orchestration frameworks~\cite{bpfd,ebpfmanager,walmartl3af} that often focus on TC/XDP programs and do not support \texttt{struct\_ops}, i.e., BPF-TCP. Similarly, while our prior work, NetEdit~\cite{meta:netedit}, provides general eBPF orchestration, \sysname extends it with the CCA-specific mechanisms described in \cref{sec:ecosystem:orchestration}.

\textbf{Transport Observability.}
Traditional TCP tools~\cite{deter,pathdump,10.5555/3691992.3692046,10.1145/2716281.2836094,dapper} diagnose application performance under well-behaved CCAs, whereas we focus on diagnosing incorrectly behaving CCAs. Other work, e.g., Web10G, externalizes limited TCP state or captures external metrics~\cite{millisampler:imc22,TcpSight}; \CCALogger instead externalizes a richer set of CCA-internal state. For example \CCALogger externalizes the following data: snd\_cwnd (a tcp\_sock’s current sending congestion window);
snd\_ssthresh (a tcp\_sock’s current slow-start size threshold);
ecn\_alpha (the fraction of sent bytes that experienced congestion in the previous RTT);
is\_loss (a flag indicating if a packet loss occurred in this RTT window);
is\_ece\_ack (a flag indicating if an ECE-marked ACK is received in this RTT window);
is\_tx\_start (a flag indicating if a transmit occurred when no packets in this RTT window);
is\_ssthresh\_invoked (a flag indicating if the hook `ssthresh` is invoked in this RTT window).

\section{Discussion}
\label{sec:discussion}

\textbf{Cloud Data Centers.} Today, we run \sysname in external cloud environments with little modification. The core issues, we find, arise from the CCAs we run on top of \sysname. In particular, a subset of our CCAs are ECN-based, and in public clouds we neither control nor have access to ECN signals. This requires us either to choose CCAs that do not need ECN or to design a new CCA that dynamically detects the available signals and selects an appropriate control loop. Today, we opt for the former and leave the latter as an open research question.

\textbf{Programmable NICs.} We focus explicitly on implementing CCAs entirely in software; however, others, e.g., Falcon~\cite{falcon:google}, have explored offloading portions of a CCA onto the NIC hardware. Fundamentally, BPF-TCP works by overriding certain kernel functionality, whereas hardware-assisted transports offload functionality to modules provided by the hardware. These approaches are not at odds with each other; consequently, \sysname could enable hardware-assisted transports in which BPF-TCP overrides some functionality and the rest is offloaded to hardware. This will likely require mechanisms that allow BPF programs to control and configure the hardware, as well as shared memory so that socket or connection state is consistently shared across both.

\textbf{Non-\texttt{struct\_ops} CCAs.} \sysname explicitly manages \texttt{struct\_ops}-based CCAs; however, others have designed CCAs using a combination of other eBPF hooks. Our core orchestrator, NetEdit, will support such CCAs, but some of \sysname's benefits, e.g., consistent storage and execution, may be harder to realize.

\textbf{Cross-CCA Fairness.} Cross-CCA fairness is traditionally managed through algorithmic design, network infrastructure configuration, or both. \sysname focuses on managing CCAs; ensuring fairness is outside the scope of the system.

\textbf{Supporting BBR.}   The reference BBRv3 implementation extends the underlying \texttt{tcp\_congestion\_ops} interface in the same direction in which we have evolved BPF-TCP. BPF-TCP exposes every \texttt{tcp\_congestion\_ops} function as a \texttt{struct\_ops} program except the diagnostic \texttt{get\_info}; this is sufficient to express BBR. BBRv3, however, adds one new callback, replaces another, and enlarges the per-connection CCA state from 104 to 144 bytes, so supporting it would require upstream interface extensions similar to our own.

\section{Conclusion and Open Research Challenges}
\label{sec:conclusion}

\sysname has allowed us to rapidly prototype, test, and deploy five eBPF-based CCAs over the last five years. Today, these CCAs carry over 97\% of our internal traffic. In this paper, we describe the evolution of our modular eBPF-based framework, the BPF-TCP hook interface, to support different CCAs, and in doing so we highlight key design lessons for modular CCAs. We also describe the challenges of maintaining application performance and reliability as we evolve our CCAs, and the challenges of extending our eBPF orchestration framework, NetEdit, to support eBPF-based CCAs. Much of our effort has focused on solutions tailored to our unique production environment, and we highlight open challenges for the research community to tackle.

\textbf{Beyond eBPF Verifier (Towards Semantic Verification):} Verifier ensure that an eBPF program does not crash the kernel; however, this is not safety. We observed crashes due to bugs in the BPF-Kernel interface and due to semantics of the program.  We encourage the research community to look into semantic verification of program behavior .
Most efforts focus on designing and engineering for single programs; however, struct\_ops are a bundle of programs that interoperate and whose correctness is a function is an emergent behavior. We similarly encourage the research community to investigate designs that look more holistically at program interactions.

\newpage
\bibliographystyle{plain}
\bibliography{ccaas}

@article{dctcp,
author = {Alizadeh, Mohammad and Greenberg, Albert and Maltz, David A. and Padhye, Jitendra and Patel, Parveen and Prabhakar, Balaji and Sengupta, Sudipta and Sridharan, Murari},
title = {Data Center TCP (DCTCP)},
year = {2010},
issue_date = {October 2010},
publisher = {Association for Computing Machinery},
address = {New York, NY, USA},
volume = {40},
number = {4},
issn = {0146-4833},
url = {https://doi.org/10.1145/1851275.1851192},
doi = {10.1145/1851275.1851192},
journal = {SIGCOMM Comput. Commun. Rev.},
month = {aug},
pages = {63–74},
numpages = {12},
}

@inproceedings {google:bolt,
author = {Serhat Arslan and Yuliang Li and Gautam Kumar and Nandita Dukkipati},
title = {Bolt: {Sub-RTT} Congestion Control for {Ultra-Low} Latency},
booktitle = {20th USENIX Symposium on Networked Systems Design and Implementation (NSDI 23)},
year = {2023},
isbn = {978-1-939133-33-5},
address = {Boston, MA},
pages = {219--236},
url = {https://www.usenix.org/conference/nsdi23/presentation/arslan},
publisher = {USENIX Association},
month = apr
}

@inproceedings{meta:netedit,
author = {Benson, Theophilus A. and Kannan, Prashanth and Gupta, Prankur and Madhavan, Balasubramanian and Arora, Kumar Saurabh and Meng, Jie and Lau, Martin and Dhamija, Abhishek and Krishnamurthy, Rajiv and Sundaresan, Srikanth and Spring, Neil and Zhang, Ying},
title = {NetEdit: An Orchestration Platform for eBPF Network Functions at Scale},
year = {2024},
isbn = {9798400706141},
publisher = {Association for Computing Machinery},
address = {New York, NY, USA},
url = {https://doi.org/10.1145/3651890.3672227},
doi = {10.1145/3651890.3672227},
booktitle = {Proceedings of the ACM SIGCOMM 2024 Conference},
pages = {721–734},
numpages = {14},
location = {Sydney, NSW, Australia},
series = {ACM SIGCOMM '24}
}

@misc{bpfd,
key={bpfd},
year = {2024},
title = {{bpfman: An eBPF Manager}},
howpublished = {\url{https://bpfman.io/main/}}
}

@ARTICLE{CTP:ToN07,
  author={Bridges, Patrick G. and Wong, Gary T. and Hiltunen, Matti and Schlichting, Richard D. and Barrick, Matthew J.},
  journal={IEEE/ACM Transactions on Networking}, 
  title={A Configurable and Extensible Transport Protocol}, 
  year={2007},
  volume={15},
  number={6},
  pages={1254-1265},
  doi={10.1109/TNET.2007.906245}}

@inproceedings{etran,
  title={$\{$eTran$\}$: Extensible Kernel Transport with $\{$eBPF$\}$},
  author={Chen, Zhongjie and Meng, Qingkai and Lao, ChonLam and Liu, Yifan and Ren, Fengyuan and Yu, Minlan and Zhou, Yang},
  booktitle={NSDI},
  year={2025}
}

@misc{ebpfmanager,
author = {{DataDog}},
year = {2023},
title = {{eBPF manager. This manager helps handle the life cycle of your eBPF programs.}},
howpublished = {\url{https://github.com/DataDog/ebpf-manager}}
}

@inproceedings{meta:ebb,
author = {Denis, Marek and Yao, Yuanjun and Hatch, Ashley and Zhang, Qin and Lim, Chiun Lin and Zhang, Shuqiang and Sugrue, Kyle and Kwok, Henry and Fernandez, Mikel Jimenez and Lapukhov, Petr and Hebbani, Sandeep and Nagarajan, Gaya and Baldonado, Omar and Gao, Lixin and Zhang, Ying},
title = {EBB: Reliable and Evolvable Express Backbone Network in Meta},
year = {2023},
isbn = {9798400702365},
publisher = {Association for Computing Machinery},
address = {New York, NY, USA},
url = {https://doi.org/10.1145/3603269.3604860},
doi = {10.1145/3603269.3604860},
booktitle = {Proceedings of the ACM SIGCOMM 2023 Conference},
pages = {346–359},
numpages = {14},
location = {New York, NY, USA},
series = {ACM SIGCOMM '23}
}

@inproceedings {metadctcp:nsdi24,
author = {Abhishek Dhamija and Balasubramanian Madhavan and Hechao Li and Jie Meng and Shrikrishna Khare and Madhavi Rao and Lawrence Brakmo and Neil Spring and Prashanth Kannan and Srikanth Sundaresan and Soudeh Ghorbani},
title = {A large-scale deployment of {DCTCP}},
booktitle = {21st USENIX Symposium on Networked Systems Design and Implementation (NSDI 24)},
year = {2024},
isbn = {978-1-939133-39-7},
address = {Santa Clara, CA},
pages = {239--252},
url = {https://www.usenix.org/conference/nsdi24/presentation/dhamija},
publisher = {USENIX Association},
month = apr
}

@ARTICLE{TcpSight,
  author={Geng, Ruopeng and Lu, Jianyuan and Fang, Chongrong and Zhang, Shaokai and Zhao, Jiangu and Zong, Zhigang and Lyu, Biao and Zhu, Shunmin and Cheng, Peng and Chen, Jiming},
  journal={IEEE Transactions on Network and Service Management}, 
  title={Enabling Stateful TCP Performance Profiling With Key Event Capturing}, 
  year={2025},
  volume={22},
  number={5},
  pages={4964-4982},
  doi={10.1109/TNSM.2025.3564336}}

@inproceedings{millisampler:imc22,
author = {Ghabashneh, Ehab and Zhao, Yimeng and Lumezanu, Cristian and Spring, Neil and Sundaresan, Srikanth and Rao, Sanjay},
title = {A Microscopic View of Bursts, Buffer Contention, and Loss in Data Centers},
year = {2022},
isbn = {9781450392594},
publisher = {Association for Computing Machinery},
address = {New York, NY, USA},
url = {https://doi.org/10.1145/3517745.3561430},
doi = {10.1145/3517745.3561430},
booktitle = {Proceedings of the 22nd ACM Internet Measurement Conference},
pages = {567–580},
numpages = {14},
location = {Nice, France},
series = {IMC '22}
}

@inproceedings{dapper,
  author       = {Mojgan Ghasemi and
                  Theophilus Benson and
                  Jennifer Rexford},
  title        = {Dapper: Data Plane Performance Diagnosis of {TCP}},
  booktitle    = {Proceedings of the Symposium on {SDN} Research, {SOSR} 2017, Santa
                  Clara, CA, USA, April 3-4, 2017},
  pages        = {61--74},
  publisher    = {{ACM}},
  year         = {2017},
  url          = {https://doi.org/10.1145/3050220.3050228},
  doi          = {10.1145/3050220.3050228},
}

@inproceedings{BPFComposer,
  author    = {Gupta, Prankur and Chahande, Takshak and Xu, Miao and
               Kannan, Prashanth and Birajdar, Tejas and Liu, Lilian and
               Krishnamurthy, Rajiv and Benson, Theophilus A.},
  title     = {{BPFComposer}: Towards Deterministic and Safe Composition of
               {eBPF} Programs (and their Shared State)},
  booktitle = {Proceedings of the 20th ACM International Conference on
               Distributed and Event-based Systems (DEBS '26)},
  year      = {2026},
  month     = jun,
  address   = {Lisbon, Portugal},
  publisher = {Association for Computing Machinery},
  pages     = {7},
  doi       = {10.1145/3809481.3812620},
  url       = {https://doi.org/10.1145/3809481.3812620},
}

@inproceedings{mtcp,
  title={$\{$mTCP$\}$: a highly scalable user-level $\{$TCP$\}$ stack for multicore systems},
  author={Jeong, EunYoung and Wood, Shinae and Jamshed, Muhammad and Jeong, Haewon and Ihm, Sunghwan and Han, Dongsu and Park, KyoungSoo},
  booktitle={11th USENIX Symposium on Networked Systems Design and Implementation (NSDI 14)},
  pages={489--502},
  year={2014}
}

@inproceedings {dctcp:morganstanley,
author = {Glenn Judd},
title = {Attaining the Promise and Avoiding the Pitfalls of {TCP} in the Datacenter},
booktitle = {12th USENIX Symposium on Networked Systems Design and Implementation (NSDI 15)},
year = {2015},
isbn = {978-1-931971-218},
address = {Oakland, CA},
pages = {145--157},
url = {https://www.usenix.org/conference/nsdi15/technical-sessions/presentation/judd},
publisher = {USENIX Association},
month = may
}

@inproceedings{tas,
  title     = {{TAS}: {TCP} acceleration as an {OS} service},
  author    = {Kaufmann, Antoine and Stamler, Tim and Peter, Simon and Sharma, Naveen Kr and Krishnamurthy, Arvind and Anderson, Thomas},
  booktitle = {European Conference on Computer Systems (EuroSys)},
  year      = {2019},
  publisher = {ACM},
}

@misc{patch3,
author    = {{Kui-Feng Lee}},
title     = {{Transit between BPF TCP congestion controls.}},
year = {2023},
howpublished = {\url{https://lore.kernel.org/all/20230323032405.3735486-1-kuifeng@meta.com/}}
}

@inproceedings{google:swift,
author = {Kumar, Gautam and Dukkipati, Nandita and Jang, Keon and Wassel, Hassan M. G. and Wu, Xian and Montazeri, Behnam and Wang, Yaogong and Springborn, Kevin and Alfeld, Christopher and Ryan, Michael and Wetherall, David and Vahdat, Amin},
title = {Swift: Delay is Simple and Effective for Congestion Control in the Datacenter},
year = {2020},
isbn = {9781450379557},
publisher = {Association for Computing Machinery},
address = {New York, NY, USA},
url = {https://doi.org/10.1145/3387514.3406591},
doi = {10.1145/3387514.3406591},
booktitle = {Proceedings of the Annual Conference of the ACM Special Interest Group on Data Communication on the Applications, Technologies, Architectures, and Protocols for Computer Communication},
pages = {514–528},
numpages = {15},
location = {Virtual Event, USA},
series = {SIGCOMM '20}
}

@inproceedings {deter,
author = {Yuliang Li and Rui Miao and Mohammad Alizadeh and Minlan Yu},
title = {{DETER}: Deterministic {TCP} Replay for Performance Diagnosis},
booktitle = {16th USENIX Symposium on Networked Systems Design and Implementation (NSDI 19)},
year = {2019},
isbn = {978-1-931971-49-2},
address = {Boston, MA},
pages = {437--452},
url = {https://www.usenix.org/conference/nsdi19/presentation/li-yuliang},
publisher = {USENIX Association},
month = feb
}

@misc{patch1,
author    = {{Martin KaFai Lau}},
title     = {{Introduce BPF STRUCT\_OPS}},
year = {2020},
howpublished = {\url{https://lore.kernel.org/bpf/20200109003453.3854769-1-kafai@fb.com/}}
}

@misc{patch4,
author    = {{Martin KaFai Lau}},
title     = {{bpf: Allow bpf tcp iter to do bpf\_(get|set)sockopt}},
year = {2021},
howpublished = {\url{https://lore.kernel.org/bpf/20210701200535.1033513-1-kafai@fb.com/}}
}

@inproceedings{snap,
  title     = {Snap: a microkernel approach to host networking},
  author    = {Marty, Michael and de Kruijf, Marc and Adriaens, Jacob and Alfeld, Christopher and Bauer, Sean and Contavalli, Carlo and Dalton, Michael and Dukkipati, Nandita and Evans, William C and Gribble, Steve and others},
  year      = {2019},
  booktitle = {Symposium on Operating Systems Principles (SOSP)},
  publisher = {ACM},
}

@misc{meta:testing,
key={meta:testing},
year = {2021},
title = {{Autonomous testing of services at scale.}},
howpublished = {\url{https://engineering.fb.com/2021/10/20/developer-tools/autonomous-testing/}}
}

@article{google:timely,
author = {Mittal, Radhika and Lam, Vinh The and Dukkipati, Nandita and Blem, Emily and Wassel, Hassan and Ghobadi, Monia and Vahdat, Amin and Wang, Yaogong and Wetherall, David and Zats, David},
title = {TIMELY: RTT-based Congestion Control for the Datacenter},
year = {2015},
issue_date = {October 2015},
publisher = {Association for Computing Machinery},
address = {New York, NY, USA},
volume = {45},
number = {4},
issn = {0146-4833},
url = {https://doi.org/10.1145/2829988.2787510},
doi = {10.1145/2829988.2787510},
journal = {SIGCOMM Comput. Commun. Rev.},
month = aug,
pages = {537–550},
numpages = {14},
}

@misc{mtp:arxiv25,
      title={A Target-Agnostic Protocol-Independent Interface for the Transport Layer}, 
      author={Pedro Mizuno and Kimiya Mohammadtaheri and Linfan Qian and Joshua Johnson and Danny Akbarzadeh and Chris Neely and Mario Baldi and Nachiket Kapre and Mina Tahmasbi Arashloo},
      year={2025},
      eprint={2509.21550},
      archivePrefix={arXiv},
      primaryClass={cs.NI},
      url={https://arxiv.org/abs/2509.21550}, 
}

@inproceedings{ccp,
  title={Restructuring endpoint congestion control},
  author={Narayan, Akshay and Cangialosi, Frank and Raghavan, Deepti and Goyal, Prateesh and Narayana, Srinivas and Mittal, Radhika and Alizadeh, Mohammad and Balakrishnan, Hari},
  booktitle={Proceedings of the 2018 Conference of the ACM Special Interest Group on Data Communication},
  pages={30--43},
  year={2018}
}

@misc{hystart++,
author    = {{P. Balasubramanian, Y. Huang, M. Olson}},
title     = {HyStart++: Modified Slow Start for TCP},
year = {2023},
howpublished = {\url{https://datatracker.ietf.org/doc/html/rfc9406}}
}

@inproceedings{bytedance:barre,
author = {Peng, Yajuan and Wei, Haoran and Zhong, Xiaolong and Huang, Junkai and Xu, Haohan and Wang, Zicheng and Bai, Yang and Jiang, Zhuo and Ye, Jianxi and Wang, Xiaoliang and Fu, Xiaoming and Dai, Huichen},
title = {Barre: empowering simplified and versatile programmable congestion control in high-speed AI clusters},
year = {2025},
isbn = {978-1-939133-48-9},
publisher = {USENIX Association},
address = {USA},
booktitle = {Proceedings of the 2025 USENIX Conference on Usenix Annual Technical Conference},
articleno = {21},
numpages = {21},
location = {Boston, MA, USA},
series = {USENIX ATC '25}
}

@misc{web10g_project,
  author       = {{Pittsburgh Supercomputing Center}},
  title        = {{Web10G:" Taking TCP Instrumentation to the Next Level}},
  howpublished = {\url{https://www.psc.edu/research/networking/web10g/}}
}

@inproceedings{falcon:google,
author = {Singhvi, Arjun and Dukkipati, Nandita and Chandra, Prashant and Wassel, Hassan M. G. and Sharma, Naveen Kr. and Rebello, Anthony and Schuh, Henry and Kumar, Praveen and Montazeri, Behnam and Bansod, Neelesh and Thomas, Sarin and Cho, Inho and Seibert, Hyojeong Lee and Wu, Baijun and Yang, Rui and Li, Yuliang and Huang, Kai and Yin, Qianwen and Agarwal, Abhishek and Vaduvatha, Srinivas and Wang, Weihuang and Moshref, Masoud and Ji, Tao and Wetherall, David and Vahdat, Amin},
title = {Falcon: A Reliable, Low Latency Hardware Transport},
year = {2025},
isbn = {9798400715242},
publisher = {Association for Computing Machinery},
address = {New York, NY, USA},
url = {https://doi.org/10.1145/3718958.3754353},
doi = {10.1145/3718958.3754353},
booktitle = {Proceedings of the ACM SIGCOMM 2025 Conference},
pages = {248–263},
numpages = {16},
location = {S{\~a}o Francisco Convent, Coimbra, Portugal},
series = {SIGCOMM '25}
}

@inproceedings {pathdump,
author = {Praveen Tammana and Rachit Agarwal and Myungjin Lee},
title = {Simplifying Datacenter Network Debugging with {PathDump}},
booktitle = {12th USENIX Symposium on Operating Systems Design and Implementation (OSDI 16)},
year = {2016},
isbn = {978-1-931971-33-1},
address = {Savannah, GA},
pages = {233--248},
url = {https://www.usenix.org/conference/osdi16/technical-sessions/presentation/tammana},
publisher = {USENIX Association},
month = nov
}

@misc{bytedance:netkit,
    author = {Chen Tang and Feng Zhou},
    title = {{Embracing Netkit Across a Million Servers, What Should We Do?}},
    howpublished = {Presented at the eBPF summit},
    year         = {2024},
    note         = {\url{https://www.youtube.com/watch?v=0w788CqTp0c&list=PLDg_GiBbAx-m7yn_FYcc41PNrgtxlISBK&index=8}}
}

@inproceedings{tran2019beyond,
  title={Beyond socket options: making the Linux {TCP} stack truly extensible},
  author={Tran, Viet-Hoang and Bonaventure, Olivier},
  booktitle={2019 IFIP Networking Conference (IFIP Networking)},
  pages={1--9},
  year={2019},
  organization={IEEE}
}

@misc{walmartl3af,
key={walmartl3af},
year = {2024},
title = {{l3af. Complete lifecycle management of eBPF programs in the kernel.}},
howpublished = {\url{https://l3af.io/}}
}

@inproceedings{10.5555/3691992.3692046,
author = {Wang, Zhe and Hu, Huanwu and Kong, Linghe and Kang, Xinlei and Ma, Teng and Xiang, Qiao and Li, Jingxuan and Lu, Yang and Song, Zhuo and Yang, Peihao and Wu, Jiejian and Yang, Yong and Ma, Tao and Liu, Zheng and Zeng, Xianlong and Cai, Dennis and Chen, Guihai},
title = {Diagnosing application-network anomalies for millions of IPs in production clouds},
year = {2024},
isbn = {978-1-939133-41-0},
publisher = {USENIX Association},
address = {USA},
booktitle = {Proceedings of the 2024 USENIX Conference on Usenix Annual Technical Conference},
articleno = {54},
numpages = {15},
location = {Santa Clara, CA, USA},
series = {USENIX ATC'24}
}

@inproceedings{10.1145/2716281.2836094,
author = {Zhou, Jianer and Wu, Qinghua and Li, Zhenyu and Uhlig, Steve and Steenkiste, Peter and Chen, Jian and Xie, Gaogang},
title = {Demystifying and mitigating TCP stalls at the server side},
year = {2015},
isbn = {9781450334129},
publisher = {Association for Computing Machinery},
address = {New York, NY, USA},
url = {https://doi.org/10.1145/2716281.2836094},
doi = {10.1145/2716281.2836094},
booktitle = {Proceedings of the 11th ACM Conference on Emerging Networking Experiments and Technologies},
articleno = {9},
numpages = {13},
location = {Heidelberg, Germany},
series = {CoNEXT '15}
}

@inproceedings {luna,
author = {Lingjun Zhu and Yifan Shen and Erci Xu and Bo Shi and Ting Fu and Shu Ma and Shuguang Chen and Zhongyu Wang and Haonan Wu and Xingyu Liao and Zhendan Yang and Zhongqing Chen and Wei Lin and Yijun Hou and Rong Liu and Chao Shi and Jiaji Zhu and Jiesheng Wu},
title = {Deploying User-space {TCP} at Cloud Scale with {LUNA}},
booktitle = {2023 USENIX Annual Technical Conference (USENIX ATC 23)},
year = {2023},
isbn = {978-1-939133-35-9},
address = {Boston, MA},
pages = {673--687},
url = {https://www.usenix.org/conference/atc23/presentation/zhu-lingjun},
publisher = {USENIX Association},
month = jul
}

@article{dcqcn,
  title={Congestion control for large-scale RDMA deployments},
  author={Zhu, Yibo and Eran, Haggai and Firestone, Daniel and Guo, Chuanxiong and Lipshteyn, Marina and Liron, Yehonatan and Padhye, Jitendra and Raindel, Shachar and Yahia, Mohamad Haj and Zhang, Ming},
  journal={ACM SIGCOMM Computer Communication Review},
  volume={45},
  number={4},
  pages={523--536},
  year={2015},
  publisher={ACM New York, NY, USA}
}

\appendix

\newpage
\textbf{Appendices are supporting material that has not been peer-reviewed.}
\newpage

\section{XRegion-CCA Synthetic Benchmark}
\label{appendix:glide:synthetic}

\begin{table}[h]
\centering
\small
\begin{tabular}{l|cc|cc|c}
\toprule
 & \multicolumn{2}{c|}{\textbf{P99 Latency}} & \multicolumn{2}{c|}{\textbf{Retransmits}} & \textbf{Thru-} \\
\textbf{Flows} & \textbf{Cubic} & \textbf{XR-CCA} & \textbf{Cubic} & \textbf{XR-CCA} & \textbf{put} \\
\midrule
12 (5MB) & 100ms & 87ms & 2.4K & 0 & +7\% \\
12 (80MB) & 765ms & 520ms & 13.3K & 0 &  \\
48 (5MB) & 275ms & 157ms & 15.7K & 1.2K & +6\% \\
48 (80MB) & 377ms & 331ms & 12.5K & 0.6K &  \\
\bottomrule
\end{tabular}
\caption{Synthetic benchmark: \Glide vs.\ Cubic with cross-region senders writing to a single receiver. Latency values are P99; throughput is aggregate improvement.}
\label{tab:glide_synthetic}
\end{table}

\begin{table*}[t]
\centering
\scriptsize
\begin{tabular}{l|ccc|cc|c}
\toprule
\textbf{Intra $\times$ Cross} & \textbf{Cross} & \textbf{Jain} & \textbf{Cross} & \textbf{Intra} & \textbf{Intra} & \textbf{Cross} \\
 & \textbf{share} & & \textbf{p50} & \textbf{p50} & \textbf{queue} & \textbf{cwnd} \\
\midrule
Cubic $\times$ Cubic        & 0.3\%  & 0.48 & 11.5~s  & 25.8~ms & 34.8~KB & 19 \\
DCTCP $\times$ Cubic        & 2.0\%  & 0.51 & 1.6~s   & 34.6~ms & 25.8~KB & 132 \\
\DTCP $\times$ Cubic        & 8.3\%  & 0.57 & 381~ms  & 37.7~ms & 21.5~KB & 579 \\
\DTCP $\times$ \Glide       & 42.3\% & 0.94 & 74.9~ms & 57.8~ms & 15.6~KB & 5098 \\
\bottomrule
\end{tabular}
\caption{Mixed-CCA coexistence. Cross share is the cross-region fraction of total
goodput; Jain is computed over all 240 flows; p50 is the median transaction time;
intra queue is the standing queue inferred from sender socket state; cross cwnd is
the mean cross-region congestion window (segments).}
\label{tab:coexistence}
\end{table*}

\newpage

\section{Mixed-CCA Coexistence}
\label{appendix:coexistence}

To quantify how our CCAs coexist, we ran a testbed experiment that retraces our
deployment history: Cubic everywhere, then DCTCP and later \DTCP for intra-region
traffic, and finally \Glide for cross-region traffic. Two cross-region senders
(${\sim}$32~ms RTT) and two intra-region senders (${\sim}$60~$\mu$s RTT) send to a
single receiver with a 400~Gbps NIC; each run uses 240 flows (60 per sender) of
8~MB request--response transactions for 600~s. Consecutive pairings change only one
side's CCA, so each difference is attributable to a single swap.
\Cref{tab:coexistence} summarizes the results.

\textbf{Fairness.} Only the production pairing (\DTCP $\times$ \Glide) avoids starving
cross-region traffic: Jain's index rises from 0.48 to 0.94. Within each class, Jain's
index stays between 0.97 and 0.99 in every run, so the unfairness is entirely between
classes.

\textbf{Queue occupancy.} \DTCP holds the lowest intra-region standing queue,
15.6~KB versus 34.8~KB under Cubic, which directly supports the buffer-headroom
effect in \cref{sec:evaluation:validation}.

\textbf{Cross-impact.} Replacing only the intra-region CCA (Cubic $\rightarrow$ DCTCP
$\rightarrow$ \DTCP) raises cross-region goodput from 0.7 to 19.9~Gbps without any
change on the cross-region side. Cubic's cross-region window never exceeds 579
segments versus 5098 for \Glide: at 32~ms RTT, the bandwidth-delay product far
exceeds what shallow ECN and drop thresholds allow, the failure that motivated
\Glide (\cref{sec:modular:xregion}). The trade-off is that intra-region traffic yields
capacity: intra-region goodput falls by 46\% and its median transaction time rises
from 25.8 to 57.8~ms, while aggregate throughput changes by only 6\%.

\textbf{Limitations.} Each pairing was run once; cross-region RTTs inflated to
48--54~ms during runs; queue occupancy is inferred from socket state rather than
switch telemetry; and the cross- and intra-region senders ran different kernel
versions.

\newpage

\section{DCNDelay-CCA Algorithm Details}
\label{appendix:d2tcp:algorithm}

\DTCP augments DCTCP's ECN response with delay-based proactive congestion window sizing. The ECN response serves as a guardrail: on each ECN mark, \DTCP includes it in the fraction of marked packets to compute the DCTCP-style alpha value, ensuring it is never more aggressive than DCTCP. The delay component computes an ideal congestion window from the observed send rate and a configured minimum RTT (75$\mu$s in-rack, 100$\mu$s in-DC, 125$\mu$s in-region):
\begin{equation}
\text{target\_cwnd} = \text{send\_rate} \times \text{min\_rtt}
\end{equation}
If the current window exceeds the target, \DTCP reduces it by 30\% of the distance to the target, a gentle reduction that avoids oscillation. When no congestion is detected, the window increases additively (as in Reno). After consecutive periods without congestion signals, the additive increase constant grows, enabling faster ramp-up on uncongested paths. On packet loss, the slow-start threshold is set to approximately 80\% of the current window ($\texttt{ssthresh} = (\texttt{cwnd} \times 819) \gg 10$).

\section{XRegion-CCA Algorithm Details}
\label{appendix:xregion:algorithm}

\Glide computes a congestion ratio after each ACK:
\begin{equation}
\texttt{ce\_ratio} = \frac{\text{\# delivered packets marked with CE}}{\text{\# total delivered packets}} \times 100\%
\end{equation}

\paragraph{Hook distribution.}
\Glide's logic is distributed across four hooks:
\begin{itemize}
    \item \texttt{pkts\_acked}: Computes the congestion ratio from ECN-marked packets and generates the continuous congestion signal that distinguishes transient from persistent congestion.
    \item \texttt{cong\_control}: Implements the INC\_CWND/DEC\_CWND state machine and the dynamic fairness threshold, bypassing the kernel's default CWR behavior.
    \item \texttt{ssthresh}: Implements Conservative Slow Start to avoid overshooting upon exiting slow start.
    \item \texttt{min\_tso\_segs}: Dynamic TSO segment override for batching efficiency.
\end{itemize}

\paragraph{Fairness via dynamic congestion thresholds.}
A static congestion threshold leads to unfairness when flows have heterogeneous RTTs: low-rate flows are more likely to exceed a fixed threshold per-ACK, causing them to back off prematurely. \Glide addresses this with a \emph{dynamic} threshold that incorporates each flow's transmission rate. A flow signals congestion when:
\begin{equation}
\texttt{tx\_rate} \times \texttt{ce\_ratio} > C
\end{equation}
where $C$ is a constant representing the targeted ECN rate. Higher-rate flows get a smaller effective threshold, causing them to react to congestion earlier and release bandwidth to lower-rate flows. This mechanism is implemented in the \texttt{cong\_control} hook, which provides the delivery-rate context needed for the computation.

\paragraph{Congestion response.}
In many ways, \Glide is an assembly of well-understood components adapted to our deployment constraints. It combines a CUBIC growth curve with DCTCP-style ECN alpha tracking, but applies gentler ECN-driven reductions ($\texttt{alpha}/3$ rather than DCTCP's $\texttt{alpha}/2$, matching BBR's ECN factor). A 6.25\% minimum floor prevents stagnation when alpha has decayed to near-zero during long uncongested periods. \Glide implements its own Proportional Rate Reduction (PRR) for smooth window reduction, bypassing the kernel's default PRR which is coupled to the CWR state machine.

The Linux kernel forces a sender into CWR upon a single ECN mark. For cross-region traffic traversing switches with shallow ECN thresholds (configured for intra-region \DTCP), this causes premature rate collapse. \Glide bypasses the kernel's congestion state machine entirely via \texttt{cong\_control} and implements its own two-state machine:

\begin{itemize}
    \item \textbf{INC\_CWND}: Increases the congestion window through three stages, exponential Slow Start ($\text{cwnd} \times 2$ per RTT), Conservative Slow Start (a HyStart++-inspired~\cite{hystart++} stage that transitions on the first ECN mark or delay increase, growing at $\text{cwnd} \times 1.05$ per RTT to avoid overshoot), and Congestion Avoidance (CUBIC growth with adaptive additive increase that grows faster when ECN is sparse). The transition from INC\_CWND to DEC\_CWND is triggered by a strong ECN signal (i.e., the dynamic threshold is exceeded) or packet loss.
    \item \textbf{DEC\_CWND}: Reduces the congestion window to a target value. For ECN-triggered events: $\text{target} = (1 - \texttt{ecn\_alpha}/3) \times \text{cwnd}$, where \texttt{ecn\_alpha} is an EWMA of \texttt{ce\_ratio}. For loss-triggered events: $\text{target} = 0.7 \times \text{cwnd}$. The sender uses \Glide's own PRR to reach the target within one RTT, then transitions back to INC\_CWND via a timer. The timer-driven transition (rather than event-driven) avoids over-reacting to transient congestion.
\end{itemize}

\section{Future Work}
\label{appendix:future:work}

\textbf{Managing Sticky Transport Signals.}
Modern CCAs rely on multiple congestion signals, some of which are negotiated or configured at different points in a connection's lifecycle. In particular some signals, notably ECN, are sticky: once negotiated during the TCP handshake, there is no standard mechanism to disable them mid-connection. This creates significant friction during CCA transitions.
For example, when rolling back from an ECN-based CCA to a non-ECN CCA (e.g., Cubic), existing connections continue to receive ECN-marked packets. Cubic interprets these marks as implicit loss signals, triggering unnecessary congestion responses and increasing retransmissions.
For the other way round, when we needed to migrate the long lived connections from non-ECN based CCA, our current mitigation in production is to transition through an intermediate CCA that gracefully handles but does not require ECN, and to use bpf\_iter to conform all active connections fleet-wide.
A cleaner solution would allow CCAs to explicitly disable or renegotiate signals on active connections. This requires both kernel support for mid-connection signal teardown and CCA-aware mechanisms to drain in-flight marked packets before transitioning. We view signal lifecycle management as a first-class concern for future modular CCA frameworks.

\textbf{Version Management and Compatibility.}
eBPF-based CCAs lack native support for versioning, creating challenges at multiple levels: tracking CCA versions, ensuring compatibility across versions, and managing shared state with non-CCA programs.

CCA versioning. When tuning parameters (e.g., alpha thresholds), adding new signals, or modifying control logic, operators face a choice: register an entirely new CCA under a different name, or overwrite the existing CCA and accept fleet-wide impact. Neither scales gracefully. Registering new CCAs proliferates names and complicates policy configuration; overwriting existing CCAs prevents incremental rollout and A/B testing. Our current workaround, encoding version information in CCA names and manually migrating connections via bpf\_iter, is operationally brittle and does not generalize.

Version compatibility. Beyond tracking versions, we lack mechanisms to express and enforce compatibility between CCA versions. When rolling out a new version, must it be backward compatible with all prior versions? With the previous N versions? How do we detect incompatibilities before they cause production impact? Currently, compatibility is validated through manual testing and operational discipline, but there are no explicit compatibility checks enforced at CCA selection or transition time. A robust solution would include version metadata in the CCA registration, compatibility matrices that the selector consults, and automated validation during rollout.

Cross-program versioning. CCA programs often share BPF maps with other programs, sockopt handlers, monitoring hooks, scope resolvers, that evolve independently. A version change in the CCA may assume a new map schema or semantic, while a dependent program still expects the old format. This expands version management from a single-program concern to an ecosystem-wide coordination problem. As discussed in \cref{sec:operational:platform}, a non-backward-compatible scope update by an unrelated binary caused widespread CCA misselection. Native support for cross-program version contracts, where shared maps declare their schema version and consumers validate compatibility, would reduce the risk of silent failures.
Future directions. We envision extending \texttt{tcp\_congestion\_ops} and the broader BPF infrastructure to support: (i) Version metadata: CCA registration includes version identifiers and feature flags, exposed to selectors and telemetry.; (ii) Compatibility declarations: CCAs declare which prior versions they can safely transition from, enabling the selector to block incompatible migrations; (iii) Shared-state contracts: BPF maps declare schema versions; programs that consume them validate compatibility at load time or runtime.
These capabilities would transform version management from an operational discipline into an enforceable system property.

\textbf{Coexistence Within the Broader eBPF Ecosystem.}
eBPF-based CCAs do not operate in isolation. A single host may run dozens of eBPF programs across diverse attach points sockops, sockopt, cgroup, tc, and struct\_ops each managed by independent teams with different priorities. Our CCA infrastructure depends on several of these: \texttt{tcp\_congestion\_ops} for the CCA itself, \texttt{bpf\_iter} for fleet-wide connection iteration, sockops for connection-level policy, and sockopt for configuration injection.
This diversity introduces subtle conflicts.  A sockops program from an unrelated team inadvertently disabled ECN by overwriting shared kernel state. The kernel provides no native ordering guarantees for cgroup-attached eBPF programs beyond insertion order, and no mechanism to express priority or mutual exclusion.
We addressed this operationally by building program chaining infrastructure that enforces priority ordering and auditing that surfaces suppressed actions. However, these are workarounds rather than solutions. Native kernel support for eBPF program composition, including explicit ordering, conflict detection, and coordinated state access, would significantly reduce operational complexity and the risk of silent failures.

\textbf{Extending to Cloud Datacenters.}
Meta is increasingly adopting cloud datacenters alongside owned infrastructure. Unlike Meta-owned facilities, cloud environments impose constraints we do not control: different network topologies, diverse hardware vendors, and varying feature support. For example, some cloud RSWs do not support ECN and rely on overprovisioning to manage congestion rendering ECN-based CCAs ineffective or counterproductive.
Onboarding custom CCAs to cloud datacenters raises several challenges: (i) Feature detection: CCAs must dynamically detect which signals (ECN, delay, loss) are meaningful in a given environment and adapt accordingly; (ii) Monitoring heterogeneity: Our current monitoring infrastructure assumes uniform CCA deployment within a region. Hybrid environments with mixed CCA capabilities require per-path or per-destination observability; (iii) Vendor coordination: Unlike owned infrastructure, we cannot patch cloud network devices. CCA designs must be robust to vendor-specific behaviors and avoid assumptions that hold only in Meta-owned networks.

We anticipate that future CCAs will need to be environment-aware, dynamically selecting signal sources and control strategies based on detected path characteristics rather than static configuration.

\end{document}